\documentclass[twoside,twocolumn,9pt]{article}
\usepackage{extsizes}
\usepackage[super,sort&compress,comma]{natbib} 
\usepackage[version=3]{mhchem}
\usepackage[left=1.5cm, right=1.5cm, top=1.785cm, bottom=2.0cm]{geometry}
\usepackage{balance}
\usepackage{mathptmx}
\usepackage{sectsty}
\usepackage{graphicx} 
\usepackage{lastpage}
\usepackage[format=plain,justification=justified,singlelinecheck=false,font={stretch=1.125,small,sf},labelfont=bf,labelsep=space]{caption}
\usepackage{float}
\usepackage{fancyhdr}
\usepackage{fnpos}
\usepackage[english]{babel}
\addto{\captionsenglish}{%
  
}
\usepackage{array}
\usepackage{droidsans}
\usepackage{charter}
\usepackage[T1]{fontenc}
\usepackage[usenames,dvipsnames]{xcolor}
\usepackage{setspace}
\usepackage[compact]{titlesec}
\usepackage{hyperref}

\usepackage{siunitx}
\DeclareSIUnit{\angstrom}{\text{\AA}}
\usepackage{booktabs}
\usepackage{xr}
\usepackage{epstopdf}

\definecolor{cream}{RGB}{222,217,201}

\begin{document}

\pagestyle{fancy}
\thispagestyle{plain}
\fancypagestyle{plain}{
\renewcommand{\headrulewidth}{0pt}
}

\makeFNbottom
\makeatletter
\renewcommand\LARGE{\@setfontsize\LARGE{15pt}{17}}
\renewcommand\Large{\@setfontsize\Large{12pt}{14}}
\renewcommand\large{\@setfontsize\large{10pt}{12}}
\renewcommand\footnotesize{\@setfontsize\footnotesize{7pt}{10}}
\makeatother

\renewcommand{\thefootnote}{\fnsymbol{footnote}}
\renewcommand\footnoterule{\vspace*{1pt}%
\color{cream}\hrule width 3.5in height 0.4pt \color{black}\vspace*{5pt}} 
\setcounter{secnumdepth}{5}

\makeatletter 
\renewcommand\@biblabel[1]{#1}            
\renewcommand\@makefntext[1]%
{\noindent\makebox[0pt][r]{\@thefnmark\,}#1}
\makeatother 
\renewcommand{\figurename}{\small{Fig.}~}
\sectionfont{\sffamily\Large}
\subsectionfont{\normalsize}
\subsubsectionfont{\bf}
\setstretch{1.125} 
\setlength{\skip\footins}{0.8cm}
\setlength{\footnotesep}{0.25cm}
\setlength{\jot}{10pt}
\titlespacing*{\section}{0pt}{4pt}{4pt}
\titlespacing*{\subsection}{0pt}{15pt}{1pt}

\fancyfoot{}
\fancyfoot[LO,RE]{\vspace{-7.1pt}\includegraphics[height=9pt]{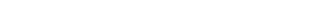}}
\fancyfoot[CO]{\vspace{-7.1pt}\hspace{11.9cm}\includegraphics{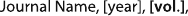}}
\fancyfoot[CE]{\vspace{-7.2pt}\hspace{-13.2cm}\includegraphics{head_foot/RF}}
\fancyfoot[RO]{\footnotesize{\sffamily{1--\pageref{LastPage} ~\textbar  \hspace{2pt}\thepage}}}
\fancyfoot[LE]{\footnotesize{\sffamily{\thepage~\textbar\hspace{4.65cm} 1--\pageref{LastPage}}}}
\fancyhead{}
\renewcommand{\headrulewidth}{0pt} 
\renewcommand{\footrulewidth}{0pt}
\setlength{\arrayrulewidth}{1pt}
\setlength{\columnsep}{6.5mm}
\setlength\bibsep{1pt}

\makeatletter 
\newlength{\figrulesep} 
\setlength{\figrulesep}{0.5\textfloatsep} 

\newcommand{\topfigrule}{\vspace*{-1pt}%
\noindent{\color{cream}\rule[-\figrulesep]{\columnwidth}{1.5pt}} }

\newcommand{\botfigrule}{\vspace*{-2pt}%
\noindent{\color{cream}\rule[\figrulesep]{\columnwidth}{1.5pt}} }

\newcommand{\dblfigrule}{\vspace*{-1pt}%
\noindent{\color{cream}\rule[-\figrulesep]{\textwidth}{1.5pt}} }

\makeatother

\twocolumn[
  \begin{@twocolumnfalse}
\vspace{1em}
\sffamily
\begin{tabular}{m{4.5cm} p{13.5cm} }

\includegraphics{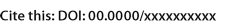} & \noindent\LARGE{\textbf{Hidden excimer formation in the gas-phase photodynamics of a BN-doped phenanthrene}}\\
\vspace{0.3cm} & \vspace{0.3cm} \\

 & \noindent\large{Jonas Fackelmayer,\textit{$^{a,\ddag}$} Michael Bühler,\textit{$^{a,\ddag}$} Michael Müller,\textit{$^{c}$} Holger Helten,\textit{$^{c}$} Ingo Fischer$^{\ast}$\textit{$^{a}$} and Merle I. S. Röhr$^{\ast}$\textit{$^{a,b}$}}\\

\includegraphics{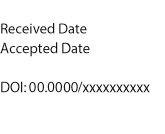} & \noindent\normalsize{Replacing \ce{CC} units by isoelectronic BN motifs provides a powerful strategy to tune the electronic structure and excited-state chemistry of polycyclic aromatic hydrocarbons (PAHs). Here, we combine multiphoton ionization spectroscopy, time-resolved photoelectron imaging, ion velocity-map imaging, and quantum-chemical calculations to disentangle the monomer and dimer photophysics of 4a,4b-azaboraphenanthrene. The monomer exhibits a structured S$_1 \leftarrow$ S$_0$ spectrum with an origin at \SI[
  uncertainty-mode = separate,multi-part-units = single]{22880(15)}{\per\centi\meter}, corresponding to \SI{2.837}{\electronvolt}, and pronounced activity in low-wavenumber deformation modes. Photoelectron spectroscopy yields an adiabatic ionization energy of \SI[uncertainty-mode = separate,multi-part-units = single
]{7.18(2)}{\electronvolt}. While the structured spectrum, high fluorescence quantum yield, small computed geometry changes, and weak spin-orbit couplings all point to a long-lived monomer S$_1$ state, time-resolved photoelectron images reveal an additional picosecond component. Ion imaging shows that this component originates from dissociative ionization of the molecular dimer, which projects dimer excited-state dynamics into the monomer mass channel. Computations identify the initially excited dimer state as a bright H-aggregate-like exciton, followed by ultrafast S$_2 \rightarrow$ S$_1$ internal conversion and subsequent structural relaxation toward an excimeric S$_1$ minimum. The experimentally observed $\approx$15 ps time constant is therefore assigned to excimer formation in the neutral dimer.}\\

\end{tabular}

 \end{@twocolumnfalse} \vspace{0.6cm}

  ]

\renewcommand*\rmdefault{bch}\normalfont\upshape
\rmfamily
\section*{}
\vspace{-1cm}


\footnotetext{\textit{$^{a}$~Institute of Physical and Theoretical Chemistry, University of Würzburg, D-97074 Würzburg, Germany; Tel: +49 931 31 89072; E-mail: ingo.fischer@uni-wuerzburg.de; merle.roehr@uni-wuerzburg.de}}
\footnotetext{\textit{$^b$} Center for Nanosystems Chemistry, University of Würzburg, D-97074 Würzburg, Germany.}
\footnotetext{\textit{$^{c}$~Institute of Inorganic Chemistry, University of Würzburg, D-97074 Würzburg, Germany. }}

\footnotetext{\dag~Supplementary Information available: [details of any supplementary information available should be included here]. See DOI: 10.1039/cXCP00000x/}

\footnotetext{\ddag~These authors contributed equally to this work.}



\section{Introduction}

The substitution of \ce{CC} units by \ce{BN} motifs in the aromatic scaffold of polycyclic aromatic hydrocarbons (PAH) is a widely explored strategy to expand the scope of organic molecules.\cite{dewar624NewHeteroaromatic1958,dewarNewHeteroaromaticCompounds1968,kawaguchiMaterialsBasedGraphite1997,liuBNCCHow2008,bosdetBNCCSubstitute2009,campbellRecentAdvancesAzaborine2012,morganEfficientSyntheticMethods2016,mcconnellLatestageFunctionalizationBNheterocycles2019,pinheiroSystematicAnalysisExcitonic2020,mullerBNBModifiedPeryleneTerrylene2025,lutzBNBNBNPhenalenyl2Deoxyuridines2026} Because \ce{BN}-doped heterocycles are isoelectronic with their all-carbon analogues, they can preserve the overall aromatic framework while displaying substantially modified electronic and optical properties. These changes arise from the polarity of the \ce{BN} bond, which introduces local dipole moments, alters the spatial distribution and character of the frontier orbitals, and changes the energetic ordering of low-lying excited states. In addition, the replacement of carbon by nitrogen (polycyclic aromatic nitrogen heterocycle, PANH) can introduce or stabilize $n\pi^*$ states, thereby opening excited-state deactivation pathways that are absent or less important in the parent hydrocarbons. 

\begin{figure}[h]
\captionsetup{justification=centering,singlelinecheck=false}
\centering
  \includegraphics[scale=0.65]{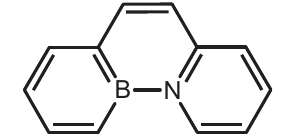}
  \caption{Structure of 4a,4b-azaboraphenanthrene $\mathbf{1}$, m/z 178/179.}
  \label{fgr:structure}
\end{figure}

As a result, \ce{BN}-doped PAHs have attracted considerable interest as chemically tunable chromophores for organic functional materials. For example, \ce{BN}-substituted polycyclic aromatic hydrocarbons have been proposed as candidates for singlet-fission (SF) materials.\cite{zengSeekingSmallMolecules2014,pinheiroSystematicAnalysisExcitonic2020,singhEnergeticsOptimalMolecular2021} As SF is a process that involves neighboring molecules, the additional interactions between individual units by dipole/dipole interactions distinguish them from the nonpolar hydrocarbons tetracene and pentacene that are known as SF materials.\cite{wilsonSingletExcitonFission2013,doverEndothermicSingletFission2018,macqueenCrystallineSiliconSolar2018,dvorakSingletFissionConcentrated2021,choiSingletFissionDynamics2023,clercqSingletFissionTIPSanthracene2024,nagayaExcitonFissionEnhanced2025} 

Therefore, not only the photophysics of the individual \ce{BN}-substituted PAH is of interest, but also the photodynamics of their dimers. In the past, photoinduced processes in dimers of anthracene, pyrene and tetracene were elucidated in the gas phase by combining time-resolved spectroscopy with high-level quantum-chemical computations.\cite{hocheMechanismExcimerFormation2017,hocheExcimerFormationDynamics2021,lemmensVibrationsPassageDiabolic2025} 

Consistent with this potential, recent years have seen growing efforts to elucidate the spectroscopy and photochemistry of \ce{BN}-substituted aromatic molecules.\cite{holzmeierPhotoionizationPyrolysis14Azaborinine2014,snyderBNDopingPhotochemistry2017,snyderExcitedStateDeactivationPathways2017,sturmImpactIsoelectronicSubstitution2024,mittagFraternalTwinsB2O22026} 
Studies on selected azaborines by infrared spectroscopy in rare-gas matrices and by gas-phase photoionization have provided valuable benchmarks, although detailed time-resolved studies of excited-state dynamics remain scarce and investigations have mainly been limited to solution.\cite{broughPhotoisomerization12Dihydro12AzaborineMatrix2012,chrostowskaUVPhotoelectronSpectroscopy122012,edel12AzaborineBoronNitrogenDerivative2015} 

Even less is known about dimers of heteroatom containing PAH.\cite{appiariusBNSubstitutionDithienylpyrenesPrevents2022,daiEclipsedTwistedExcimers2024} Azaphenanthrene-dimers have been investigated recently,\cite{miaoStackingFavoredHydrogen2022} but information on dimers of \ce{BN}-substituted PAH is not available. Recently, we investigated 4a,8a-azaboranaphthalene in the gas phase by picosecond time-resolved photoionization and found a strong decrease of the S$_1$ excited state lifetime with vibrational excitation.\cite{sturmImpactIsoelectronicSubstitution2024} The observed time-dependence was rationalized by a conical intersection, separated from the S$_1$ state by a barrier that can be surmounted by vibrational excitation. Extending such studies to larger \ce{BN}-doped PAHs raises an additional question: whether the polar \ce{BN} unit not only modifies monomer relaxation, but also creates aggregate-specific excited-state pathways such as exciton relaxation and excimer formation.

Here, we address this question for 4a,4b-azaboraphenanthrene \textbf{1}, a \ce{BN}-doped phenanthrene in which the \ce{BN} unit is located at the 4a,4b bay position (see Fig.~\ref{fgr:structure}). This substitution pattern preserves the overall phenanthrene topology while introducing a polarized \ce{BN} unit into the central ring. 

Experimentally, the \ce{BN} dopant induces a pronounced increase in photoluminescence efficiency relative to the undoped phenanthrene. Photoluminescence quantum yields of 0.64 in THF and 0.58 in cyclohexane were reported, compared with 0.09 for phenanthrene, accompanied by an emission shift to shorter wavelengths in the blue spectral region. Consistently, fluorescence lifetimes of 5.3 ns in cyclohexane and 4.1 ns in THF were measured, highlighting this class of \ce{BN}-incorporated aromatic compounds as promising candidates for blue-emitting fluorophores.\cite{bosdetBlueFluorescent4aAza4bboraphenanthrenes2007,mullerBNPhenanthreneBNPyreneBasedFluorescent2024} These properties make \textbf{1} an instructive system in which a bright and apparently long-lived monomer chromophore can be compared with the excited-state dynamics of its weakly bound dimer. Here, we investigate the excited-state processes in isolated \textbf{1} and its dimer in a free jet using picosecond time-resolved photoionization spectroscopy, with the goal to investigate it without perturbation by a solvent and compare it directly with electronic-structure calculations. As probe methods, we chose both photoion and photoelectron detection. The latter, applied by us in the form of photoelectron imaging (TR-PEI, time-resolved photoelectron imaging), is known to be particularly sensitive to changes in the electronic states during deactivation. Numerous reviews describe a range of applications of time-resolved photoelectron spectroscopy (TR-PES) and TR-PEI.\cite{stolowFemtosecondTimeResolvedPhotoelectron2004,roderExploringExcitedStateDynamics2019,schuurmanTimeresolvedPhotoelectronSpectroscopy2022} 

To put the effects of replacing \ce{CC} by \ce{BN}-units into context, results on \textbf{1} are compared with those for the parent PAH phenanthrene and for azaphenanthrenes. Phenanthrene has been extensively characterized in solution and in the gas phase.\cite{offenFluorescenceLifetimesAromatic1968,warrenVibronicActivityLaser1986,karcherSpectralAtlasPolycyclic1988,salamaElectronicAbsorptionSpectroscopy1994,schmittElectronicContinuaTimeresolved2001,hanspiestVibrationalSpectroscopyGasphase2001,salamaPOLYCYCLICAROMATICHYDROCARBONS2011,kowakaElectronicVibrationalRotational2012,gehmDeterminationRelativeIonization2018,nazariUltrafastDynamicsPolycyclic2019}  Spectra of the S$_1$ $\leftarrow$ S$_0$ transition are well resolved, while for the S$_2$  $\leftarrow$ S$_0$ transition lifetimes decrease rapidly with excitation.\cite{schmittElectronicContinuaTimeresolved2001,stolowFemtosecondTimeResolvedPhotoelectron2003} High-level quantum-chemical calculations provide a robust description of its electronic structure and photophysics.\cite{gonzalez-luqueTheoreticalCharacterizationAbsorption2003,abengozarNewMemberBNPhenanthrene2019,zhangElectronDelocalizationLowlying2023} 

Resonance-enhanced multiphoton ionization (REMPI) spectra of azaphenanthrenes also show well resolved vibrational structure around the origin, but lifetimes start to decrease at around $\approx \SI{+1600}{\per\centi\meter}$.\cite{sturmElectronicStructuresAzaphenanthrenes2024}  By time-resolved X-ray spectroscopy it was found that the deactivation involves low-lying n$\pi^*$ states. \cite{schaffnerTimeresolvedXraySpectroscopy2025} 
As will be shown below, the properties of the \ce{BN}-substituted PAH \textbf{1} differs quite a bit from the one of simple PAH or PANH. 

\section{Methods}

The experimental setup has been described in detail previously.\cite{flockTimeresolvedPhotoelectronImaging2019,auerswaldElectronicStructurePyracene2013} It consists of a picosecond \SI{10}{Hz} laser system from Ekspla and a gas-phase time-of-flight (TOF) and velocity-map imaging (VMI) spectrometer. Briefly, the \SI{351}{nm} 3rd harmonic of a Nd:YLF laser (generated from \SI{85}{\percent} of the fundamental output at \SI{1053}{nm}, yielding \SI{5}{mJ}) was coupled into an optical parametric generator (OPG) producing tunable pump pulses (\SIrange{0.7}{1.1}{\milli\joule}). For the probe pulse, the 4th harmonic at \SI{263.25}{nm} (\SI{\sim 25}{\micro \joule}) was employed. The instrument response function (IRF) was around \SI{4}{ps}, with a spectral bandwidth of  roughly \SI{25}{\per\centi\meter}.
The experiments were conducted in a differentially pumped vacuum chamber and the sample was heated to 128~$^\circ$C, seeded in Argon ($p_0$(Ar) = \SI{1.2}{bar}) and expanded through a solenoid pulsed valve with a \SI{0.3}{mm} diameter nozzle into the vacuum. 

4a,4b-Azaboraphenanthrene was synthesized according to the route established by Bosdet \textit{et al.}\cite{bosdetBlueFluorescent4aAza4bboraphenanthrenes2007} The purity was checked using $^1$H-NMR and $^{11}$B\{$^{1}$H\}-NMR spectroscopy, also air stability for at least two weeks could be verified, see Electronic Supplementary Information (ESI).

The REMPI spectrum was obtained by averaging three wavelength scans. In each scan the OPG was tuned in \SI{0.1}{nm} steps and each data point was averaged over 50 laser shots. The laser beam was weakly focused (f = \SI{1000}{mm}) into the ionization region. 

To study the changes in the excited-state lifetimes, several vibronic bands were investigated in a [1+1$^\prime$]-process with the \SI{263.25}{nm} probe pulse. The time-resolved spectra were averaged over \numrange{10}{30} time-delay scans. In each scan, a single data point was averaged over 50 laser shots. The time intervals between the points varied between \SI{0.5}{ps} close to the pump–probe overlap, and up to some \SIrange{10}{100}{ps} far away from it. All decay traces were fitted employing a biexponential decay. 

In addition, photoelectron images were recorded for a number of S$_1$ bands, which yielded the ionization energy (IE) of the molecule. For the imaging photoelectron experiments each 2D raw image was averaged over 10000 shots for static spectra and 2000 -- 4000 for all time-resolved measurements. The images were reconstructed employing the pBASEX\cite{garciaTwodimensionalChargedParticle2004} algorithm up to second order Legendre polynomial. 

Quantum-chemical calculations aimed at elucidating the vibrational spectrum were conducted within the framework of (time-dependent) Density Functional Theory (TD-DFT). The calculations employed the $\mathrm{\omega}$B97X-D functional\cite{chaiLongrangeCorrectedHybrid2008} along with the augmented correlation-consistent polarized triple-zeta (aug-cc-pVTZ) basis set.\cite{kendallElectronAffinitiesFirstrow1992} The computations were executed using the Gaussian 16 quantum-chemical software package.\cite{m.j.frischGaussian16Revision2016} The choice of a large basis set augmented with diffuse functions was imperative for the accurate description of low-lying Rydberg states in TD-DFT calculations. 

The molecular structures were optimized, and Hessian matrices were computed for both the ground (S$_0$) and the first excited (S$_1$) states. Vibrationally resolved absorption spectra for the S$_1 \leftarrow \text{S}_0$ transition were simulated using the Herzberg-Teller approximation. This was achieved using a time-independent approach at a temperature of 50 K, employing the adiabatic Hessian (AH) model as implemented in the FCclasses3.0 program.\cite{santoroEffectiveMethodCompute2007,santoroEffectiveMethodComputation2008,fabriziosantoroandjaviercerezoFCclasses32019} To characterize the relevant electronic transitions, the Natural Transition Orbitals (NTO) formalism was employed.\cite{martinNaturalTransitionOrbitals2003} 

The dimer structures were initially generated using CREST~3.0\cite{prachtCRESTProgramExploration2024} with GFN2-xTB,\cite{bannwarthGFN2xTBAnAccurateBroadly2019} as implemented in the xTB program.\cite{bannwarthExtendedTightbindingQuantum2021} They were subsequent refined with ORCA 6.0.1\cite{neeseSoftwareUpdateORCA2025} at the $\omega$B97X-D3/aug-cc-pVDZ level of theory,\cite{kendallElectronAffinitiesFirstrow1992,linLongRangeCorrectedHybrid2013} employing the tighter numerical integration grids provided by the \texttt{DefGrid3} setting. A detailed description of the conformational sampling and subsequent refinement procedure is provided in the ESI.

\section{Results}
\subsection{Characterization of structure and electronic properties}
\begin{figure}
    \centering
    \includegraphics[]{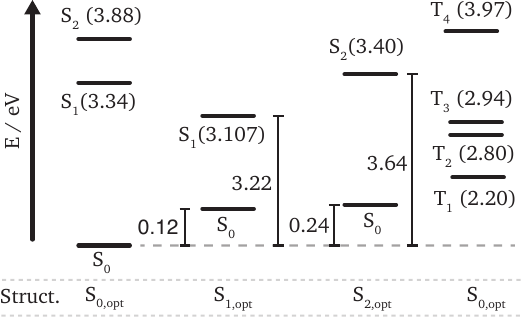}
    \caption{Potential energy scheme of the low-lying singlet and triplet states of structure \textbf{1}, computed at the $\omega$B97X-D/aug-cc-pVTZ level of theory. The S$_0$, S$_1$ and S$_2$ energies are shown at the corresponding optimized singlet state geometries, allowing comparison of vertical and adiabatic excitation energies. Triplet state energies T$_1$ to T$_4$ are reported at the S$_0$-optimized geometry.}
    \label{fig:energy_scheme}
\end{figure}
To assess the optically accessible excited states, the vertical excitation energies of the low-lying singlet states were computed and are summarized in Tab. \ref{tab:excited_states}. The S$_1$ state at $E_{\mathrm{vert}} = \SI{3.34}{\electronvolt}$ exhibits the largest oscillator strength ($f_\text{osc}=0.29$), indicating that this transition will dominate the UV/Vis absorption spectrum. The next higher states are characterized by substantially lower oscillator strengths.
\begin{table}[b]
  \centering
  \caption{Calculated vertical electronic excitation energies $E_{\text{vert}}$,
  wavelengths $\lambda$ and oscillator strengths $f_\text{osc}$ for the singlet
  excited states of the monomer.}
  \label{tab:excited_states}
  \begin{tabular}{lccc}
    \toprule
    Electronic state & $E_{\text{vert}}$/eV & $\lambda$/nm & $f_\text{osc}$ \\
    \midrule
    S$_1$ & 3.34  & 371.55 & 0.29 \\
    S$_2$ & 3.88 & 319.65 & 0.03 \\
    S$_3$ & 4.25 & 291.56 & 0.09 \\
    S$_4$ & 5.11 & 242.85 & 0.20 \\
    \bottomrule
  \end{tabular}
\end{table}
The optimized geometries for the monomer structure of the S$_0$ and S$_1$ states, as well as that of the cationic ground state D$_0$, are shown in Fig. \ref{SI-fig:struct_monomer}. The calculated structural changes are generally small, with the largest difference observed for the \ce{BC} bond length in the central ring, which increases from 1.51 \r{A} in S$_0$ to 1.54 \r{A} in S$_1$.

The energy scheme in Fig.~\ref{fig:energy_scheme} summarizes the geometry-dependent excitation energies, including the vertical and adiabatic transitions as well as the triplet-state energies. As visible, the adiabatic excitation energy E$_{\text{ad}}$ of the S\textsubscript{1} state is \SI{3.22}{\electronvolt} (indicated as the energy above the S$_0$ equilibrium structure), thus there is a slight geometry change upon excitation. The S$_2$ state is slightly higher in energy with $\mathrm{E}_{\mathrm{ad}} = \SI{3.64}{\electronvolt}$. From the natural transition orbitals, given in Fig. \ref{SI-fig:NTOs_S1S2}, it is concluded that the S$_1$ to S$_4$ states all have a $\pi\pi^*$ character. 

It is noteworthy that the computations yield three triplet states with energies below the S$_1$ state, all of them with $\pi\pi^*$ character (see Fig. \ref{SI-fig:NTO_Triplets}). As intersystem crossing (ISC) is a potential deactivation pathway, spin-orbit coupling coefficients (SOC) from S$_1$ into the various triplet states were computed. The values are summarized in Fig. \ref{SI-fig:SOC}, Tabs. \ref{SI-tab:SOC_opt} and \ref{SI-tab:ISC_T}. As expected for an El-Sayed forbidden $^1\pi\pi^* \leftarrow ^3\pi\pi^*$ ISC, all values are below \SI{0.5}{\per\centi\meter}, which is comparably small.\cite{el-sayedSpinOrbitCoupling1963} 

\subsection{Energy-resolved spectra}

A time-of-flight mass spectrum exhibits practically only the peaks at m/z 178 and 179, corresponding to the $^{10}$B and $^{11}$B isotopologues of the molecule. In addition, contributions of the $^{13}$C isotopologue are visible.

\begin{figure}[t]
\centering
  \includegraphics{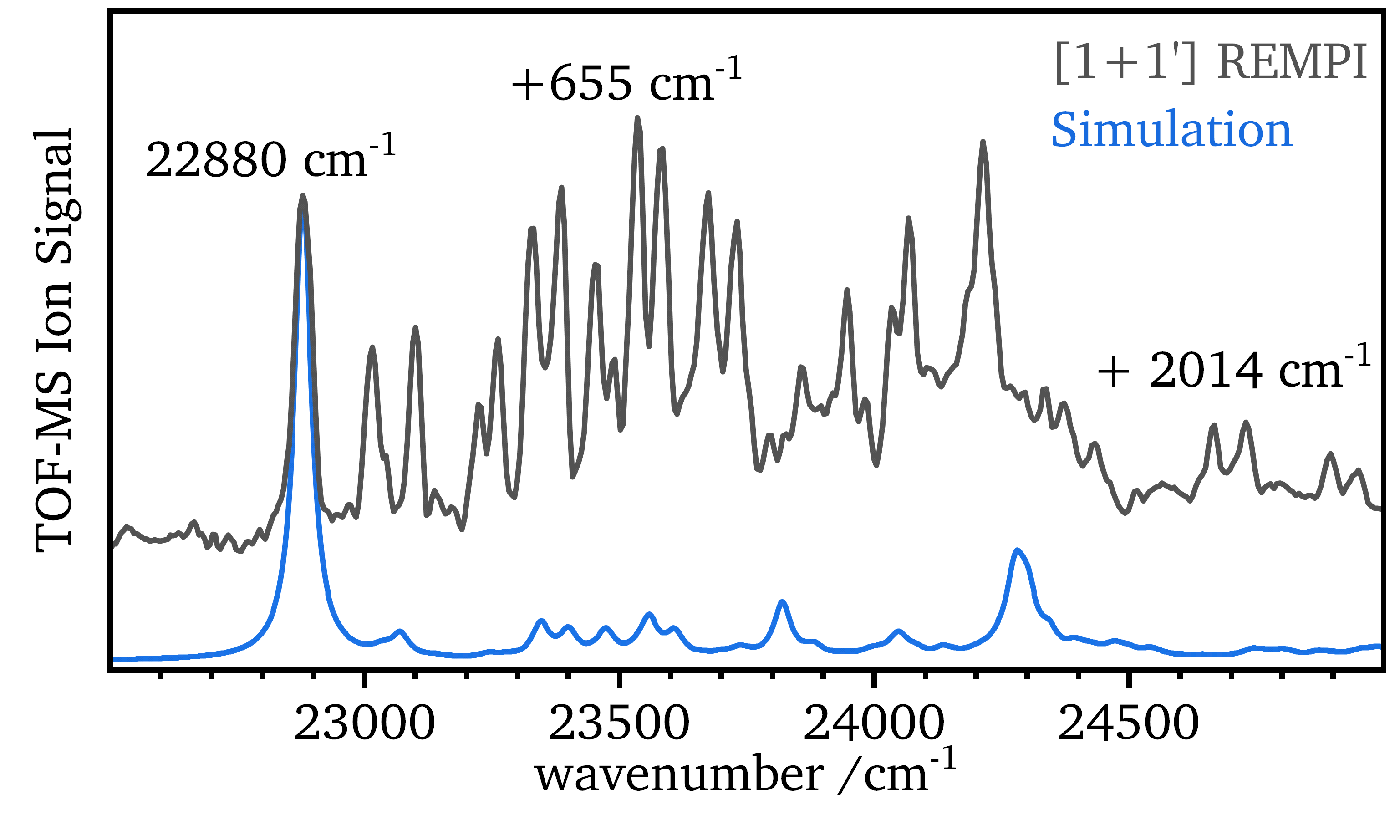}
  \caption{[1+1$^\prime$] REMPI spectrum of compound \textbf{1} (grey) with a DFT simulation of the spectrum (blue).}
  \label{fgr:REMPI}
\end{figure}

Fig. \ref{fgr:REMPI} shows the two-color REMPI of the molecule.  As visible, the spectrum shows vibrational resolution in the low energy part. At higher energies bands begin to overlap. Several of these vibronic bands have a considerable intensity. 

By comparison with the liquid-phase spectrum \cite{mullerBNPhenanthreneBNPyreneBasedFluorescent2024} and computations (cf. Tab.~\ref{tab:excited_states} and Figs.~\ref{SI-fig:abs_fluor} and~\ref{SI-fig:calc_vs_exp}), the transition is assigned to the S\textsubscript{1}~$\leftarrow$~S\textsubscript{0} transition. The band at \SI[uncertainty-mode = separate,multi-part-units = single]{22880 \pm 15}{\per\centi\meter} (\SI[uncertainty-mode = separate,multi-part-units = single]{2.837 \pm 0.002}{\eV}) is assigned to the origin of this transition. The error is estimated from the full width at half maximum (FWHM) of the band. Its E$_{\text{ad}}$ is thus \SI{0.38}{\electronvolt} lower than calculated, a value that is within the range of error of DFT computations. Interestingly, the S\textsubscript{1}~$\leftarrow$~S\textsubscript{0} transition is lowered by around \SI{0.8}{\electronvolt} compared to the parent PAH phenanthrene. 

The blue trace in Fig. \ref{fgr:REMPI} represents a simulation. While the band positions are represented reasonably well, the simulated intensities are significantly lower. In particular, the activity in the low-frequency bending modes is significantly underestimated, which indicates that the geometry change is underestimated in the computations. Two rather intense bands are visible at +135 and \SI{+220}{\per\centi\meter} and possibly appear in combination with the most intense band at \SI{+655}{\per\centi\meter}. This band is assigned to a mode associated with an increase of the  BN and BC bond lengths in the excited state. Due to the laser bandwidth, transitions at higher energy overlap and are difficult to assign. The computed and observed vibronic transitions are listed and compared in Tabs. \ref{SI-tab:freq_S0_S1_D0} and \ref{SI-tab:VibExpVCalc} in the ESI.

\begin{figure}[]
\centering
  \includegraphics{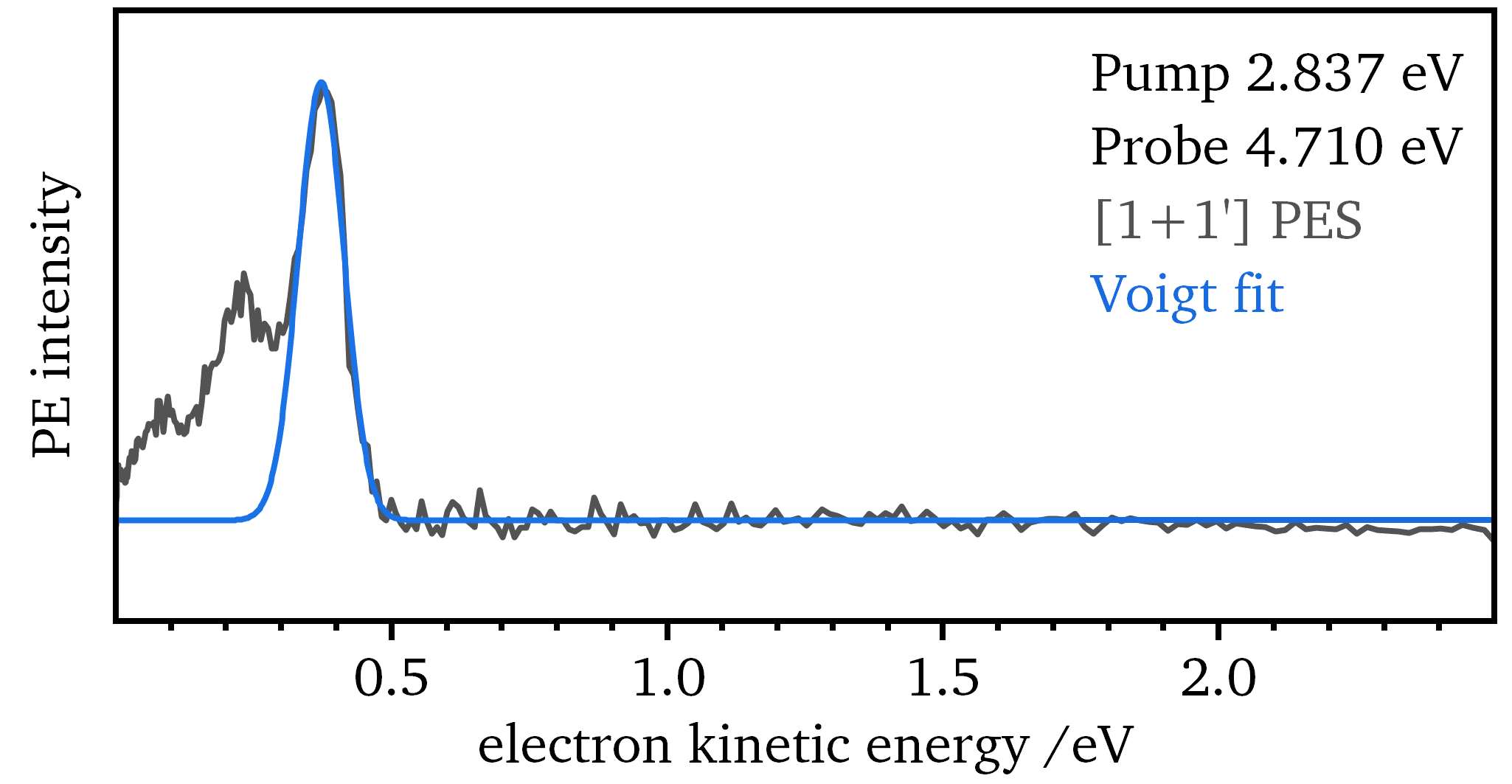}
  \caption{[1+1$^\prime$] photoelectron spectrum recorded via the S\textsubscript{1} origin. The signal onset was fitted with a 
  Voigt profile. An IE of \SI[uncertainty-mode = separate,multi-part-units = single]{7.18\pm 0.02}{\eV} was determined from the center of the Voigt profile.}
  \label{fgr:IE}
\end{figure}

A photoelectron spectrum recorded via the 0\textsuperscript{0} transition is shown in Fig. \ref{fgr:IE}. As visible, there is a pronounced onset at an electron kinetic energy (eKE) of \SI{0.5}{\electronvolt}, which is fitted by a Voigt profile. From this fit, an adiabatic ionization energy IE$_{\text{ad}}$ of \SI[uncertainty-mode = separate,multi-part-units = single]{7.18\pm 0.02}{\eV} is extracted, using the expression
\begin{equation}
\text{IE}_{\text{ad}}=h\nu - \text{eKE}  
\end{equation}
The experimentally determined IE$_\text{ad}$ is in good agreement with the computed value of \SI{6.91}{\eV}. In comparison, for the PAH parent phenanthrene an $\mathrm{IE}_{\mathrm{ad}} =$ \SI{7.89}{\electronvolt} has been evaluated,\cite{hagerTwolaserPhotoionizationSupersonic1988} thus the IE is significantly decreased upon inserting \ce{BN}. In contrast, for 4a,8a-azaboranaphthalene an $\mathrm{IE}_{\mathrm{ad}} =$ \SI{8.27}{\electronvolt} has been found, slightly higher than the value of \SI{8.14}{\electronvolt} reported for naphthalene. \cite{sturmImpactIsoelectronicSubstitution2024,cockettVibronicCouplingGround1993} This shows that the energetics strongly depends on the exact position of the \ce{BN} unit in the molecule. 

In Fig. \ref{fgr:IE}, some vibrational activity is visible at low eKE. The energy difference $\Delta\mathrm{eKE} \approx \SI{0.15}{\electronvolt}$ could match transitions into bending modes of the cation. As visible in Fig. \ref{SI-fig:struct_monomer}, the computed geometry change going from S$_1$ to D$_0$ is small, thus diagonal Franck--Condon factors (FCF) are found in simulations and the vibrational activity is surprising from a theory point of view. We will discuss an alternative explanation below.

\subsection{Time-delay scans}

In the next series of experiments, we recorded the ion signal as a function of time at selected excitation energies. We chose the S$_1$ origin and two bands at \SI{+665}{\per\centi\meter} and \SI{+2014}{\per\centi\meter} for time-resolved experiments. The results are summarized in Fig. \ref{SI-fig:IonTraces}. 
The time constant of the ion signal increases with higher excitation energies from $\approx \SI{10}{\pico\second}$ for the 0\textsuperscript{0} transition to $\approx \SI{35}{\pico\second}$ for the transition at \SI{+2014}{\per\centi\meter}.

Furthermore, a considerable offset remains in the ion signal. This shows that the final state of the deactivation can be efficiently ionized under the experimental conditions. However, a picosecond transient is at least for low-lying vibronic levels in contradiction with the previously reported high fluorescence quantum yield.\cite{bosdetBlueFluorescent4aAza4bboraphenanthrenes2007,mullerBNPhenanthreneBNPyreneBasedFluorescent2024}

\begin{figure}[]
\centering
      \includegraphics{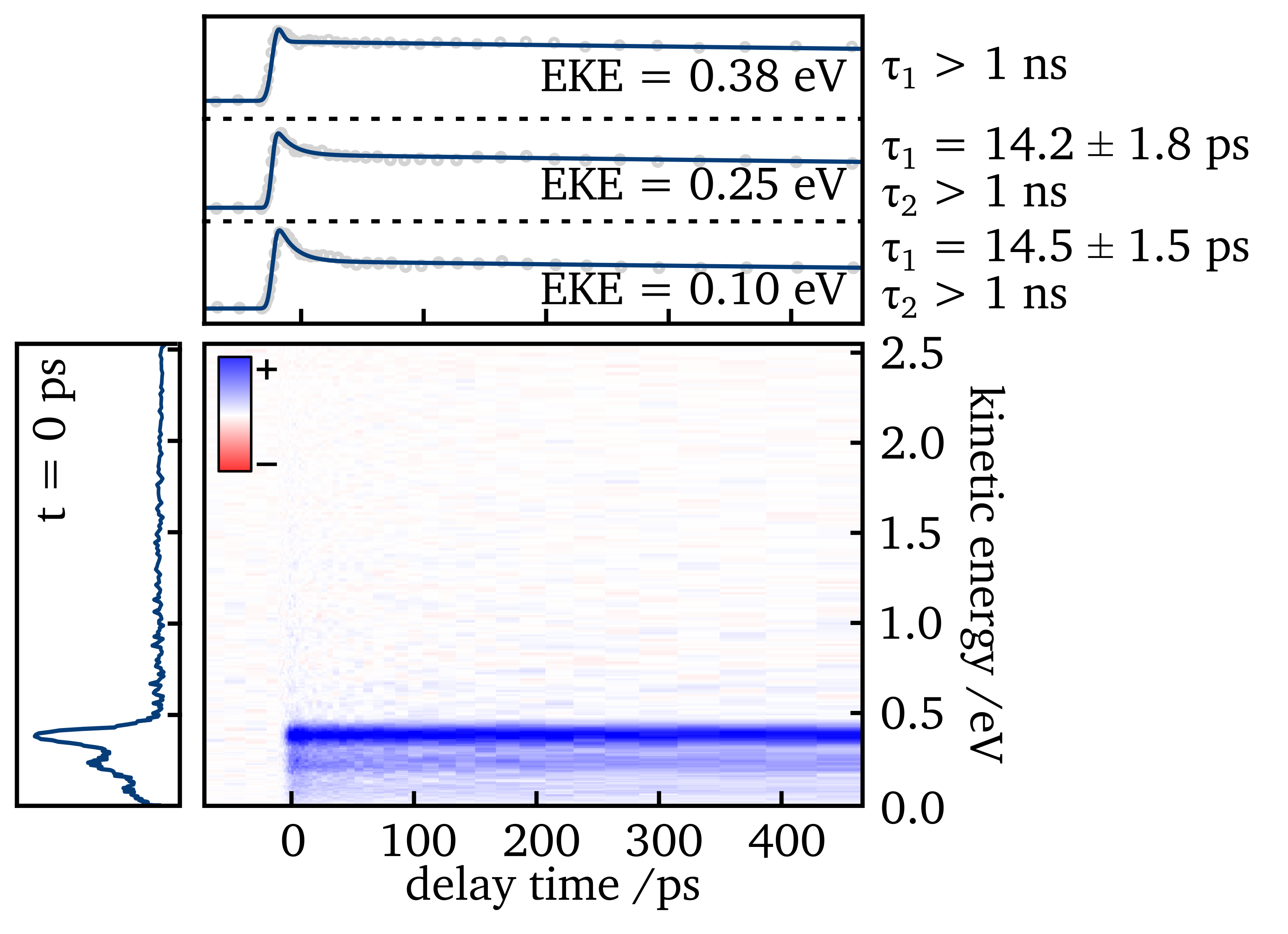}
  \caption{Time-resolved map of the photoelectron signal, recorded upon excitation of the S$_1$ origin. Top: Delay traces of electrons at specified kinetic energies. Left: PES at t$_0$.}
  \label{fgr:TRPES}
\end{figure}

As has been shown by several groups, photoelectron detection can yield additional information on the excited-state dynamics, because it can distinguish transitions into different final states of the cation.\cite{stolowFemtosecondTimeResolvedPhotoelectron2004,roderExploringExcitedStateDynamics2019,schuurmanTimeresolvedPhotoelectronSpectroscopy2022} 
Fig. \ref{fgr:TRPES} displays a 2D map, plotting the photoelectron signal as a function of time delay and of eKE. The map was recorded by exciting the S$_1$ origin, i.e. at a pump energy of \SI{2.837}{\electronvolt}. The time-independent background signals due to the pump or probe laser alone have been subtracted. Red regions in the 2D map indicate a signal decrease upon interaction with the pump laser, while the blue color indicates an increase. The 2D map is dominated by transitions from the S$_1$ 0\textsuperscript{0} state into the ionic ground state at low eKE as discussed above. Energy conservation confirms that the signal is due to a [1+1$^\prime$] process. 

 On top of the 2D map, the time dependence of the bands at $\mathrm{eKE} = \SI{0.38}{\electronvolt}$, corresponding to the vibrational ground state, and at \SI{0.25}{\electronvolt} as well as \SI{0.1}{\electronvolt} are shown. Interestingly, the time dependence is different. The most intense band at \SI{0.38}{\electronvolt} shows  no decay within the investigated delay range, apart from a minute contribution that corresponds to the instrument response function (IRF) and is due to additional multiphoton contributions around t\textsubscript{0}. The fit suggests a time constant on the ns scale. This is in agreement with the observed strong fluorescence from the S$_1$ state. However, the low energy bands at \SI{0.25}{\electronvolt} and \SI{0.1}{\electronvolt} show a weak decay with a time constant $\tau \approx \SI{15}{\pico\second}$,  comparable to the decay visible in the ion signal, cf. Fig. \ref{SI-fig:IonTraces}. 
 
 As the measured time-dependence should not depend on the vibrational state of the ion, one might conclude that it is due to vibrationally excited states populated in sequence band transition from vibrationally excited S$_0$. However, the expected diagonal FCF for photoionization due to the small geometry change rule out this explanation. 

\begin{figure}[t]
\centering
\includegraphics{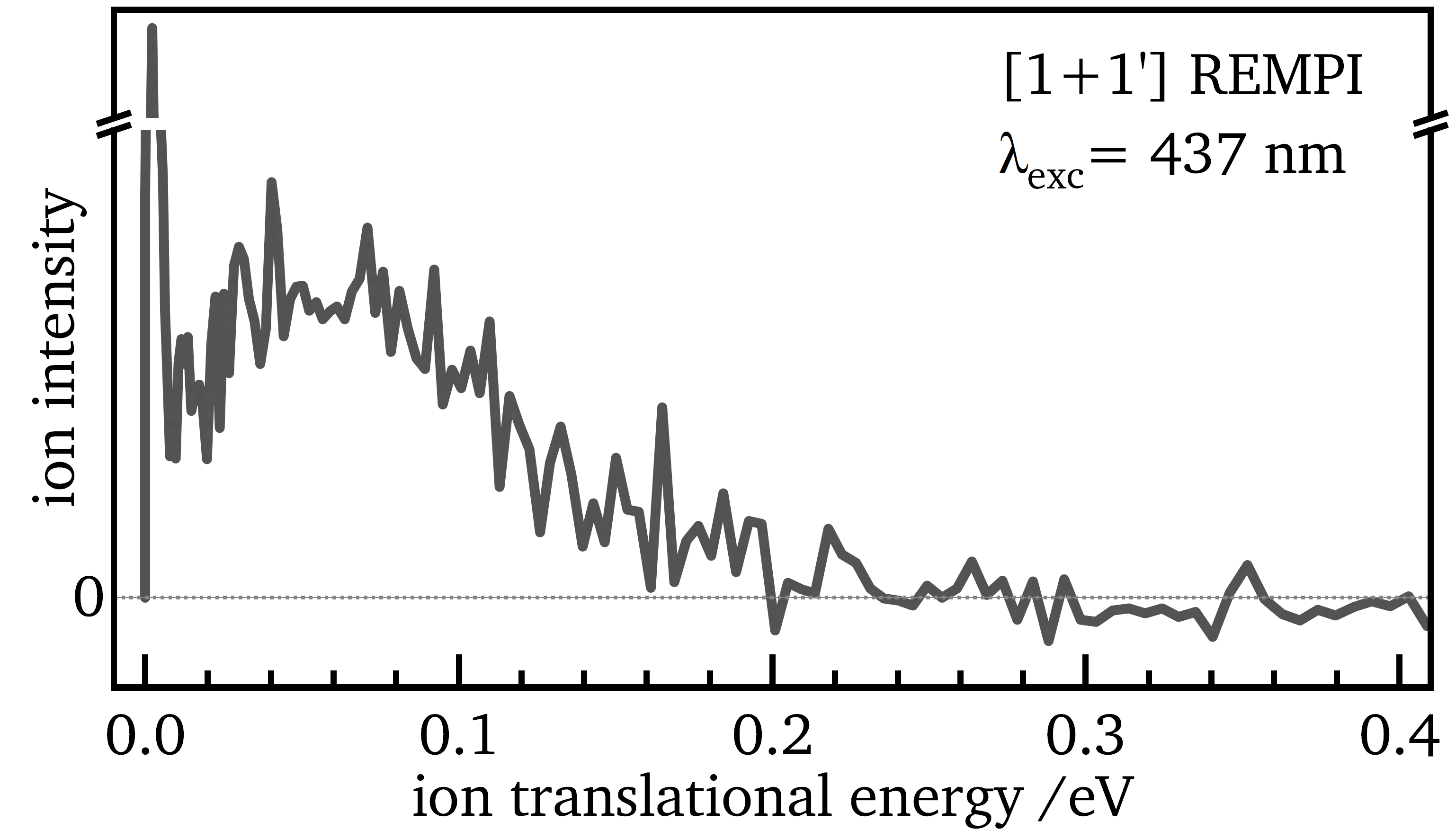}
  \caption{Translational kinetic energy release of the monomer cations at t$_0$.}
  \label{fgr:VMI}
\end{figure}

\begin{figure}[]
\centering
  \includegraphics{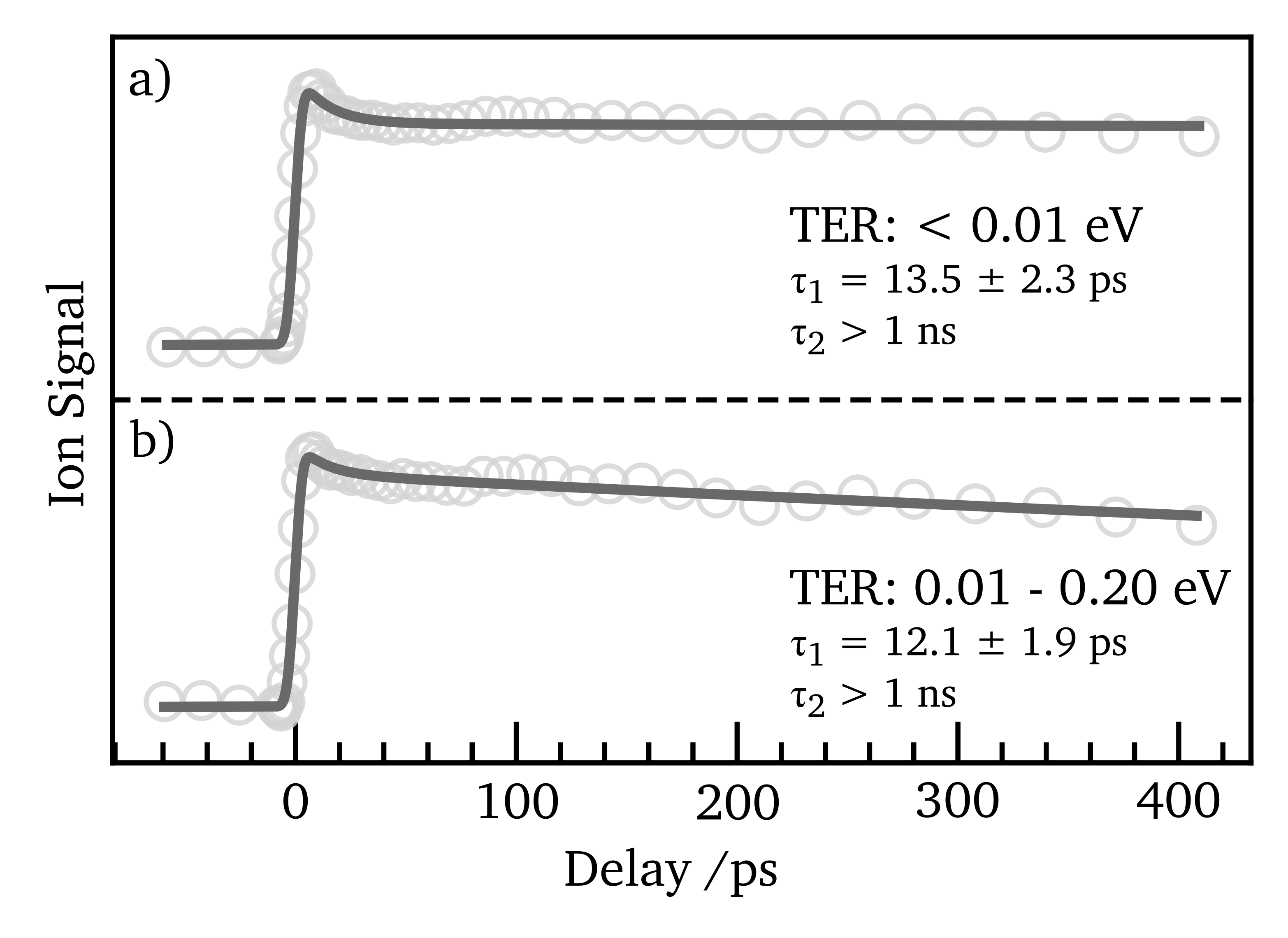}
  \caption{Comparison between monomer ion signal for different translational kinetic energy releases as extracted from the VMI ion signal.}
  \label{fgr:VMI-time}
\end{figure}
Given that the molecule is (a) strongly fluorescent in solution \cite{bosdetBlueFluorescent4aAza4bboraphenanthrenes2007}, (b) the vibrational structure in the photoelectron spectrum is in contrast to the expected diagonal FCF for ionization, (c) intersystem crossing can be excluded based on the computed SOC coefficients (see Fig. \ref{SI-fig:SOC}) and (d) a change in laser polarization yielded very similar spectra, excluding rotational effects, we  had to explore alternative explanations for the dynamics observed in the TR-PES (see Fig.~\ref{fgr:TRPES}). We therefore examined, whether fragments from dissociation of clusters contribute to the molecular signal of \textbf{1}, despite the minuscule dimer signal in the mass spectrum (see Fig. \ref{SI-fig:Tof_Dimer}). 

Velocity map imaging (VMI) of the ion signal provides the opportunity to identify such contributions. Fig.~\ref{fgr:VMI}, recorded at the S$_1$ origin (\SI{22880}{\per\centi\meter}), shows the monomer signal as a function of the ion translational energy (TER, translational energy release). A sharp peak at very low translational energy is due to jet cooled monomer ions. In addition, a rather broad distribution is visible that extends to 200 meV. Such a broad distribution is indicative of a signal contribution from preceding fragmentation in the ion.

Fig. \ref{fgr:VMI-time} shows the time-dependence at a) low TER close to zero and b) at translational energies between \SI{0.01}{\electronvolt} and \SI{0.20}{\electronvolt}. As visible, a decay with $\tau \approx \SI{13}{\pico\second}$ and small amplitude is apparent. The signal at $\mathrm{TER} < \SI{0.01}{\electronvolt}$ shows the same behavior as the one at higher TER, so its dynamics also corresponds to the fragment contribution that extends to very low energies. These data strongly suggest that the time-dependence is due to fragment ions from the dissociation of clusters, most likely the dimer of \textbf{1}.

\subsection{Structural characterization of the dimer}
\begin{figure}
    \centering
    \includegraphics[]{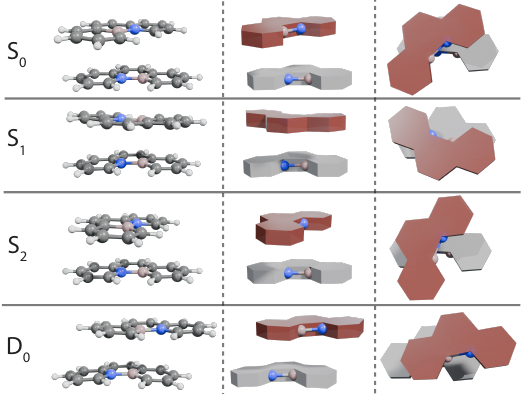}
    \caption{Optimized dimer structures of the S$_0$, S$_1$, S$_2$, and D$_0$ states. Simplified side and bird's-eye-view representations are included to illustrate the relative orientation of the dimer units.}
    \label{fig:dimer_structs}
\end{figure}

This observation motivated a computational examination of the dimer, starting with a characterization of the most stable dimer geometries. The calculations yield two low-energy dimer configurations within \SI{0.05}{\eV} that might contribute to the signal. 

\begin{figure}
    \centering
    \includegraphics[]{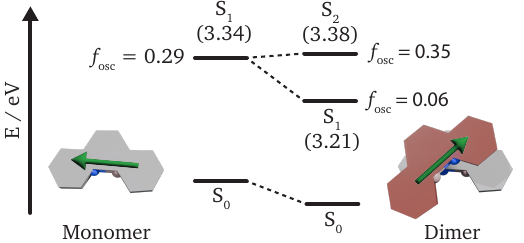}
    \caption{Excitonic splitting of the lowest electronic transition upon formation of the dimer. The monomer transition at 3.34~eV with an oscillator strength of $f_{\mathrm{osc}} = 0.29$ splits into two dimer exciton states at 3.21 and 3.38~eV with oscillator strengths of $f_{\mathrm{osc}} = 0.06$ and $f_{\mathrm{osc}} = 0.35$, respectively. The redistribution of oscillator strength to the higher-energy exciton state is characteristic of H-aggregate like coupling. Dark green arrows indicate the monomer transition dipole moments obtained from transition charges and their relative orientation in the dimer geometry.}
    \label{fig:Dimer_H_Agg}
\end{figure}

In the global minimum structure of the electronic ground state (Fig. \ref{fig:dimer_structs}, top trace), the $\pi$-systems of the two monomers are arranged approximately parallel to one another and are slightly rotated within the molecular plane (see Fig.~\ref{fig:dimer_structs}, center and right column). Due to the parallel alignment of the transition dipole moments, which lie within the $\pi$-plane, the lowest-energy dimer geometry can be described as H-aggregate-like. Consistent with this arrangement, the S$_2$ state is the bright excitonic state, whereas the S$_1$ state is darker and exhibits a lower oscillator strength (see Figs. \ref{fig:Dimer_H_Agg} and Tab.~\ref{tab:dimer_exc_osc}). The other identified optimized dimer structures also indicate H-aggregate-type coupling (see Fig. \ref{SI-fig:dimer_tddft}). 

\begin{table}[htbp]
\centering
\caption{Calculated vertical electronic excitation energies $E_{\mathrm{vert}}$, wavelengths $\lambda$ and oscillator strengths $f_\text{osc}$ for the singlet excited states of the ground-state-optimized dimer.}
\label{tab:dimer_exc_osc}
\begin{tabular}{c c c c}
\toprule
Electronic state & $E_{\mathrm{vert}}$/eV & $\lambda$/nm & $f_\text{osc}$ \\
\midrule
S$_{1}$ & 3.21 & 386.70 & 0.06 \\
S$_{2}$ & 3.38 & 366.60 & 0.35 \\
S$_{3}$ & 4.02 & 308.80 & 0.00 \\
S$_{4}$ & 4.05 & 306.40 & 0.00 \\
\bottomrule
\end{tabular}
\end{table}
For the excited states, a distinctly different picture is obtained. To describe the relative arrangement of the two monomer units, we use a set of geometric parameters collected in Tab.~\ref{tab:bn-lowest-energy-plane-curvature-tilt}. Here, $d_{\mathrm{p-p}}$ denotes the distance between the two fitted monomer planes, and $\alpha_{\mathrm{p}}$ is the angle between these planes. The quantities $\bar{d}{\mathrm{N-p}}$ and $\bar{d}{\mathrm{B-p}}$ measure the average distance of the N and B atoms from the opposite monomer plane, respectively, and therefore capture local contacts that can differ from the average plane--plane distance when the molecules are curved. The parameter $\kappa_{\mathrm{rms}}$ quantifies the root-mean-square curvature of the monomer framework; larger values indicate a stronger deviation from planarity. The energetically preferred S$_1$ structure consists of two antiparallel $\pi$ surfaces. These surfaces are almost parallel to each other and show the smallest interplane tilt among the BN-substituted structures, with $\alpha_{\mathrm{p}} = 1.72^\circ$. At the same time, the molecular framework is more strongly curved than in S$_0$, as reflected by an increase of $\kappa_{\mathrm{rms}}$ from 0.007 to 0.012 \text{\AA}$^{-1}$. The distance between the two surfaces is reduced from 3.38 \r{A} in the ground-state geometry to 3.17 \r{A} in the optimized S$_1$ minimum (see Tab. \ref{tab:bn-lowest-energy-plane-curvature-tilt}), indicating an enhanced attractive interaction in the excited state. 
\begin{table}[htbp]
\centering
\caption{Distances, curvature, and interplane angle for the lowest-energy BN dimer structures; p denotes the fitted monomer plane and {$\alpha_{\mathrm{p}}$}\ is the interplane tilt angle. Detailed definitions are given in the ESI.}
\label{tab:bn-lowest-energy-plane-curvature-tilt}
\begin{tabular}{cccccc}
\toprule
\shortstack{Electronic\\state} & \raisebox{0.5\normalbaselineskip}{$d_{\mathrm{p}\text{-}\mathrm{p}}$/\text{\AA}} & \raisebox{0.5\normalbaselineskip}{$\bar{d}_{\mathrm{N}\text{-}\mathrm{p}}$/\text{\AA}} & \raisebox{0.5\normalbaselineskip}{$\bar{d}_{\mathrm{B}\text{-}\mathrm{p}}$/\text{\AA}} & \raisebox{0.5\normalbaselineskip}{$\kappa_{\mathrm{rms}}$/\text{\AA}$^{-1}$} & \raisebox{0.5\normalbaselineskip}{$\alpha_{\mathrm{p}}$/deg} \\
\midrule
S$_0$ & 3.38 & 3.38 & 3.30 & 0.007 & 6.99 \\
S$_1$ & 3.17 & 3.15 & 3.09 & 0.012 & 1.72 \\
S$_2$ & 3.25 & 3.28 & 3.16 & 0.005 & 7.65 \\
D$_0$ & 3.22 & 3.14 & 3.08 & 0.024 & 4.91 \\
\bottomrule
\end{tabular}
\end{table}
Owing to this curved arrangement, the local atom-to-plane separations are even shorter than the average interplanar distance, with mean \ce{N}-to-plane and \ce{B}-to-plane distances of 3.15 \r{A} and 3.09 \r{A}, respectively.

\begin{figure}
    \centering
    \includegraphics[]{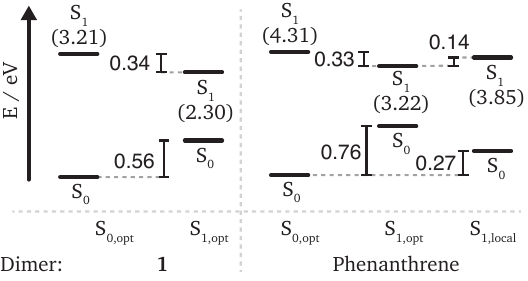}
    \caption{Ground-state energy differences between the minimum S\(_0\)-optimized and S\(_1\)-optimized geometries of \textbf{1} and phenanthrene dimer. For phenanthrene dimer, the S\(_{1,\mathrm{local}}\) structure, which most closely resembles the S\(_1\)-optimized geometry of \textbf{1} dimer, is additionally shown. This local S\(_1\) minimum lies \SI{0.14}{eV} above the global S\(_1\) minimum of phenanthrene. Values in parentheses denote the corresponding S\(_1\) excitation energies. All energies were calculated at the \(\omega\)B97X-D3/aug-cc-pVDZ level of theory.}
    \label{fig:dimer_opt_energy}
\end{figure}

This behavior differs from that of the corresponding phenanthrene reference structures (see Tab.~\ref{SI-tab:phenanthrene-lowest-energy-plane-curvature-tilt} and Fig.~\ref{SI-fig:CC_all}). In phenanthrene, the S$_1$-optimized structure retaining an orientation similar to the S$_0$ minimum is a local minimum that lies 0.14~eV above the global S$_1$ minimum (see Fig. \ref{fig:dimer_opt_energy}). Its plane--plane distance decreases from 3.44 to 3.24~\AA{} upon S$_1$ optimization. However, the two fitted planes are more strongly tilted, with $\alpha_{\mathrm{p}} = 8.83^\circ$, while both local C--plane distances remain at 3.29~\AA{}. Thus, the phenanthrene dimer contracts in a more displaced and tilted arrangement of its $\pi$ surfaces, whereas the BN-substituted dimer forms a nearly parallel contact with locally shortened N- and B-to-plane distances. Nevertheless, relaxation to the global S$_1$ minimum yields similar stabilization energies of 0.34~eV for \textbf{1} and 0.33~eV for phenanthrene dimer (see Fig.~\ref{SI-fig:energy_cc}).

For \textbf{1}, optimization on the S$_1$ surface leads to more than one minimum. The structure obtained when starting from the global S$_0$ minimum remains closer to the ground-state arrangement and lies more than {$0.1 \mathrm{eV}$} above the lowest S$_1$ minimum. 

\begin{figure}
    \centering
    \includegraphics[]{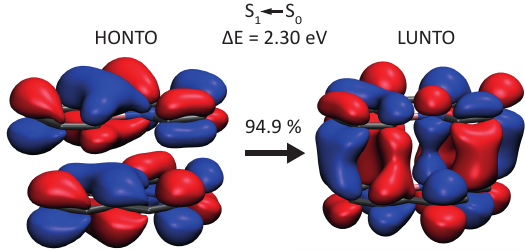}
\caption{Natural transition orbitals of the optimized S$_1$ minimum-energy structure. The dominant $\mathrm{HONTO} \rightarrow \mathrm{LUNTO}$ transition accounts for $94.9\,\%$ of the S$_1 \leftarrow \text{S}_0$ excitation with $\Delta E = 2.30\,\mathrm{eV}$. The $\mathrm{LUNTO}$ displays a strongly delocalized character over the dimer structure.}
\label{fig:S1_NTOs}
\end{figure}

The NTO analysis of this lowest S$_1$ structure (see Fig.~\ref{fig:S1_NTOs}) supports its assignment as an excimer-like state. The transition is dominated by a $\mathrm{HONTO} \rightarrow \mathrm{LUNTO}$ contribution of $94.9\%$ at an excitation energy of 2.30~eV. While the $\mathrm{HONTO}$ is mainly localized on one monomer, the $\mathrm{LUNTO}$ extends over both units. This redistribution of the excited electron density, together with the shorter intermolecular distance, the nearly parallel arrangement of the two $\pi$ surfaces, and the increased curvature of the monomer frameworks, is characteristic of an excimeric S$_1$ minimum.

The S$_2$ state shows a less pronounced structural reorganization relative to S$_0$. Its interplane distance is only moderately reduced, and the two monomer planes remain more tilted than in the relaxed S$_1$ minimum. We therefore expect the Franck--Condon overlap from the S$_0$ geometry to be more favorable for excitation into S$_2$ than into the strongly reorganized S$_1$ minimum. This geometric effect reinforces the oscillator-strength redistribution discussed above: the upper exciton component is both brighter and more readily accessed from the ground-state dimer geometry.

In the case of the cation (Fig.~\ref{fig:dimer_structs}, bottom trace), the two molecular $\pi$ planes are again oriented approximately parallel to each other but are laterally displaced such that the two \ce{B} atoms are positioned nearly directly above one another.
Similar to the optimized S$_1$ structure, the $\pi$ surfaces exhibit a pronounced curvature. This curvature is even larger in the cationic state, with $\kappa_{\mathrm{rms}} = 0.024$ \text{\AA}$^{-1}$, compared to 0.012 \text{\AA}$^{-1}$ for S$_1$. Consequently, the mean \ce{B}-to-plane distance is reduced to 3.08 \r{A}, although the corresponding interplanar distance remains larger at 3.22 \r{A}. The mean \ce{N}-to-plane distance is similarly shortened to 3.14 \r{A}. This difference reflects the curved arrangement of the molecular framework, which brings the heteroatoms closer to the opposite molecular plane than suggested by the average interplanar separation. Ionization from S$_1$ and S$_2$ is therefore associated with a considerable geometry change, and transitions into low-lying bound vibrational levels of the ion are expected to have small FCFs.

The structural comparison suggests a specific role of the \ce{BN} unit in the dimer. In the S$_1$ minimum, the antiparallel \ce{BN} arrangement allows the two $\pi$ systems to approach each other in an almost parallel geometry, while the molecular frameworks bend slightly toward one another. As a result, the local \ce{N}- and \ce{B}-to-plane contacts become shorter than the average interplane distance alone would suggest. The same tendency toward curved, locally shortened contacts is also found in the cationic state. This geometry change is important for the interpretation of the experiment, because it explains why ionization of the relaxed dimer is expected to have poor Franck--Condon overlap with low-lying bound ionic levels and can instead lead efficiently to dissociative photoionization.

\subsection{Deactivation pathway of the dimer}
In contrast to the monomer, photoexcitation of the dimer gives rise to a substantially more complex excited-state landscape. While the monomer undergoes only minor structural relaxation upon excitation, the dimer supports additional configurations and electronically coupled excitonic states, which open alternative stabilization and deactivation pathways. 

\begin{figure}[]
    \centering
    \includegraphics[]{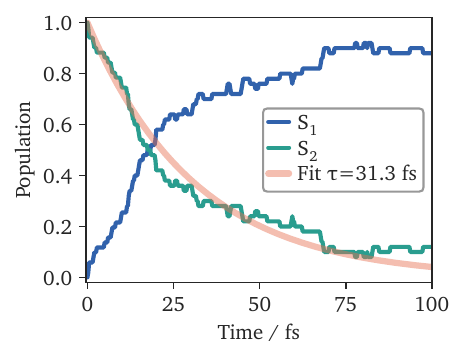}
    \caption{Population dynamics of the dimer initiated in S$_2$ from 50 Wigner sampled initial conditions at the S$_0$-optimized geometry. The S$_2$ decay was fitted with a single exponential function, yielding $\tau = 31.3$ fs and indicating ultrafast population transfer to S$_1$.}
    \label{fig:S2_dynamic}
\end{figure}

To assess whether the optically bright excitonic S$_2$ state can efficiently couple to S$_1$, trajectory-based nonadiabatic dynamics simulations were performed. In total, 50 trajectories were propagated for 100 fs at the TDDFT level of theory, with computational details provided in the ESI. Since the main objective was to characterize the early-time S$_2$-to-S$_1$ population transfer, this simulation window is sufficient to resolve the relevant nonadiabatic process. The S$_2$ decay was quantified by fitting the time-dependent S$_2$ population with a monoexponential function, yielding a time constant of 31.3 fs (see Fig. \ref{fig:S2_dynamic}). 

\begin{figure}
    \centering
    \includegraphics[]{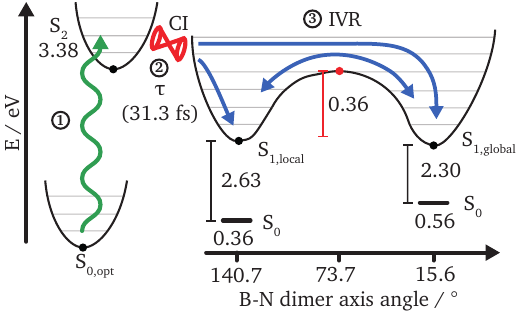}
    \caption{Schematic representation of the \ce{BN} dimer exciton relaxation pathway. Step \textbf{1} indicates excitation from S$_0$ into the bright S$_2$ state, followed by rapid relaxation to S$_1$ within $\tau = 31.3$ fs in step \textbf{2}. Step \textbf{3} describes IVR on the S$_1$ surface, which enables relaxation into local and global minima. The global minimum represents the most stable excimer state. The local-to-global S$_1$ barrier of 0.36~eV, obtained using the nudged elastic band (NEB) method (see Fig. \ref{SI-fig:NEB_path}), is associated with rotation of the \ce{BN} planes.\cite{asgeirssonNudgedElasticBand2021} The \ce{BN} dimer axis angle is defined such that 0$^\circ$ corresponds to an aligned \ce{BN-BN} configuration.}
    \label{fig:Dimer_path}
\end{figure}

The resulting mechanism can be described as a three-step relaxation pathway (see Fig. \ref{fig:Dimer_path}). First, optical excitation populates the bright S$_2$ state of the dimer. Second, the system undergoes ultrafast internal conversion through an S$_2$/S$_1$ conical intersection, leading to efficient population transfer to S$_1$ with a fitted time constant of 31 fs, which cannot be resolved experimentally. This rapid decay indicates strong nonadiabatic coupling between the initially populated excitonic state and the lower-lying S$_1$ state. Third, the molecule reaches S$_1$ in a vibrationally hot configuration, from which further relaxation proceeds by intramolecular vibrational redistribution (IVR) and structural reorganization on a longer timescale. 

Along this relaxation coordinate, both an S$_1$ local minimum, which remains structurally similar to the S$_0$ equilibrium geometry, and the S$_1$ global minimum are accessible. These two minima differ markedly in the relative orientation of the two monomer units, as reflected by the \ce{B-N}/\ce{B-N} dimer-axis angle, which changes from 140.7° in the local S$_1$ minimum to 15.6° in the global S$_1$ minimum. The latter exhibits pronounced excimer character and is stabilized by a substantially lowered S$_1$ energy. The barrier connecting the local and global S$_1$ minima was evaluated using the nudged elastic band method and amounts to 0.36~eV, placing it below the available vibrational excess energy after S$_2$ to S$_1$ conversion (see Fig.~\ref{SI-fig:NEB_path}). Consequently, access to the excimer-like global minimum is energetically feasible, especially under experimental conditions where higher vibrational levels of S$_2$ are initially populated. This provides a consistent mechanistic picture in which bright excitonic absorption is followed by ultrafast S$_2$ to S$_1$ internal conversion and subsequent vibrational cooling toward either locally excited or excimer-stabilized S$_1$ minima.

\section{Discussion}
The most intriguing observation in the data is a time dependence in the TR-PES (see Fig.~\ref{fgr:TRPES}) that cannot be rationalized by the properties of \textbf{1}. Instead, the ion VMI, cf. Figs~\ref{fgr:VMI} and \ref{fgr:VMI-time}, shows that the observed decay is most likely due to fragment ions originating from the dimer of \textbf{1}. The obvious presence of dimers prompts the question, why only minuscule amounts are observed in the mass spectra. The explanation is provided by the computed geometries. The global minimum in S$_0$ corresponds to a structure that resembles an H-aggregate, i.e. both rings are parallel, although rotated by 180° and the twisting angle is small. For the S$_2$ state, which corresponds to the upper component of the exciton splitting, a similar arrangement was found. Together with the high oscillator strength, this ensures a bright and efficient S$_2$ $\leftarrow$ S$_0$ transition. 

In the ion however, the minimum-energy geometry corresponds to a structure with both rings significantly displaced with respect to each other. As a consequence, Franck--Condon factors are small and ionization will preferentially terminate in excited states of the ion that are possibly (pre-)dissociative.  Given the computed energetics and the large geometry change upon ionization, it is not surprising that dissociative ionization of the dimer dominates. 

The  adiabatic ionization energy of the dimer is computed to be reduced to \SI{6.45}{\electronvolt} and the onset for dissociative photoionization calculated to be \SI{7.55}{\electronvolt} (see Fig. S20 in the ESI). Based on the maximum TER of \SI{0.18}{\electronvolt} obtained from Fig.~\ref{fgr:VMI} and a photon energy of \SI{7.55}{\electronvolt} deposited in a [1+1$^\prime$] process, dissociative photoionization sets in between 7.35 and \SI{7.40}{\electronvolt}, in good agreement with the computed value. Despite the energetic accessibility of the bound cation, dissociative states are efficiently accessed upon photoionization because the corresponding Franck--Condon factors increase with photon energy. Thus, the time-dependence of the fragment ions carries the signature of the dynamics of the dimer.

The wavelength dependence of the high-energy part of the TER is given in the ESI, Fig. \ref{SI-fig:REMPI_Dimer}. As visible, it is broad and unstructured, in agreement with the computed short lifetime, so excitation of the S$_1$ origin will necessarily be accompanied by excitation of the dimer. Due to the small geometry change upon ionization, photoionization of \textbf{1} from the S$_1$ origin will terminate in the vibrational ground state and thus be associated with high eKE.  Photoionization of the dimer on the other hand will terminate in excited vibrational states of the ion and emit photoelectrons with low eKE, because of the large geometry change. Therefore, the highest energy photoelectrons in Fig. \ref{fgr:TRPES} are associated with the monomer and show no fast time-dependence, whereas the low-energy electrons correspond to the dimer of \textbf{1}.

The population dynamics indicate an ultrafast population transfer from S$_2$ to S$_1$ with a time constant of 31 fs. This time constant cannot be resolved within the experiment, due to the limited temporal resolution of the ps laser setup (IRF$\approx$ 4 ps). However, the S$_1$ structure close to the conical intersection differs strongly from the global minimum structure of the S$_1$ state. Therefore, intramolecular vibrational energy redistribution (IVR) will set in, associated with structural relaxation, as indicated in Fig.~\ref{fig:Dimer_path}. We therefore
assign $\tau_1$ to geometric relaxation within the S$_1$ manifold. Based on the computed strong decrease of the inter-ring distance, this relaxation is associated with excimer formation. 

The photoionization probability decreases, because FCF for transitions to the ion become smaller. Nevertheless, there is considerable excess energy deposited in S$_1$, which allows access to large parts of the potential energy surface and thus a significant ionization probability at all times. In previous work on pyrene and tetracene, time constants of 6 ps and 62 ps were found for excimer formation, in both cases in very good agreement with theory.\cite{hocheMechanismExcimerFormation2017,hocheExcimerFormationDynamics2021} The time constant of $\SI{\approx15}{\pico\second}$ found here is in a similar range.  

\section{Conclusion}

The origin of the S$_1 \leftarrow$ S$_0$ transition is found at \SI{22880}{\per\centi\meter}, corresponding to a red shift of approximately 0.8 eV relative to phenanthrene. Photoelectron spectroscopy yields an adiabatic ionization energy of \SI[
  uncertainty-mode = separate,multi-part-units = single]{7.18(2)}{\electronvolt}. TD-DFT calculations show that the low-lying singlet and triplet states of the monomer are all of $\pi\pi^*$ character, with small geometry changes upon excitation and ionization. Consistent with this picture, the dominant photoelectron signal associated with the monomer S$_1$ state persists on the nanosecond time scale.

In addition to this long-lived monomer contribution, however, time-resolved photoelectron spectra reveal a weaker picosecond component with a time constant of $\tau \approx \SIrange{15}{20}{\pico\second}$. Ion velocity-map imaging shows that this time dependence originates from the dimer of \textbf{1}, whose dissociative ionization contributes to the monomer mass channel. Quantum-chemical calculations rationalize this observation. The neutral dimer adopts an H-aggregate-like ground-state structure, for which the upper exciton component corresponds to the bright optically accessible state. Upon excitation, this S$_2$ state decays to S$_1$ within 31 fs, a process that is too fast to be resolved with the present picosecond experiment. The subsequent relaxation on the S$_1$ surface involves structural reorganization of the dimer, including a reduction of the inter-plane distance from 3.38 \r{A} in the electronic ground state to 3.17~\r{A} in the relaxed excited-state structure. This relaxation leads to the formation of an excimer-like S$_1$ minimum and is assigned to the experimentally observed picosecond component.

The combined experimental and computational analysis therefore shows that 4a,4b-azaboraphenanthrene behaves as a bright and comparatively long-lived monomer, while its dimer opens a hidden excimer-forming relaxation channel. Because ionization of the dimer is accompanied by substantial geometry change and dissociation, the dimer dynamics becomes visible only indirectly through the monomer mass channel. These results demonstrate that \ce{BN} substitution does not merely shift the monomer spectrum, but can also control intramolecular excited-state relaxation and excimer formation in weakly bound PAH aggregates.
\section*{Author Contributions}
JF: Investigation (spectroscopy), formal analysis, data curation, validation, visualization, writing – original draft, writing – review \& editing, conceptualization; MB: Investigation (theory), writing – original draft, writing – review \& editing, conceptualization, methodology, validation, visualization; MM: Investigation (synthesis); IF: conceptualization, supervision, writing – original draft, writing – review \& editing, project administration, funding acquisition; MISR, HH: conceptualization, supervision, writing – review \& editing, project administration, funding acquisition.

\section*{Conflicts of interest}
The authors declare no conflicts of interest.

\section*{Data availability}

The data underlying this study are openly available in the GitHub repository at \href{https://github.com/roehr-lab/4a_4b_Azaboraphenanthrene}{\url{https://github.com/roehr-lab/4a_4b_Azaboraphenanthrene}}. The deposited experimental data include TOF mass, REMPI, and photoelectron spectra, translational energy distributions of electrons and ions, delay trace datasets, and maps from the time resolved photoelectron measurements. The computational data comprise Cartesian coordinates of the monomer structures in the S$_0$, S$_1$, S$_2$, and D$_0$ states and of the investigated dimer structures, as well as the data for the frequency resolved and ensemble spectra, ISC rates, SOC values, surface hopping trajectories, and counterpoise correction data for the dissociation energy.

\section*{Acknowledgements}
The work has been financially supported by the Deutsche Forschungsgemeinschaft (DFG, German Research Foundation) – Project-ID 551403841 – SFB 1762 “Boron as Property-Determining Element”. 


\bibliography{BN_Phen} 
\bibliographystyle{rsc} 

\end{document}


\pagestyle{plain}
\thispagestyle{plain}
\fancypagestyle{plain}{
\renewcommand{\headrulewidth}{0pt}
}

\makeFNbottom
\makeatletter
\renewcommand\LARGE{\@setfontsize\LARGE{15pt}{17}}
\renewcommand\Large{\@setfontsize\Large{12pt}{14}}
\renewcommand\large{\@setfontsize\large{10pt}{12}}
\renewcommand\footnotesize{\@setfontsize\footnotesize{7pt}{10}}
\makeatother

\renewcommand{\thefootnote}{\fnsymbol{footnote}}
\renewcommand\footnoterule{\vspace*{1pt}%
\color{cream}\hrule width 3.5in height 0.4pt \color{black}\vspace*{5pt}} 
\setcounter{secnumdepth}{5}

\makeatletter 
\renewcommand\@biblabel[1]{#1}            
\renewcommand\@makefntext[1]%
{\noindent\makebox[0pt][r]{\@thefnmark\,}#1}
\makeatother 
\renewcommand{\figurename}{\small{Fig.}~}
\sectionfont{\sffamily\Large}
\subsectionfont{\normalsize}
\subsubsectionfont{\bf}
\setstretch{1.125} 
\setlength{\skip\footins}{0.8cm}
\setlength{\footnotesep}{0.25cm}
\setlength{\jot}{10pt}
\titlespacing*{\section}{5pt}{4pt}{20pt}
\titlespacing*{\subsection}{5pt}{15pt}{1pt}
\titleformat{\section}{\huge\bfseries}{\thesection}{1em}{}          


\renewcommand{\headrulewidth}{0pt} 
\renewcommand{\footrulewidth}{0pt}
\setlength{\arrayrulewidth}{1pt}
\setlength{\columnsep}{6.5mm}
\setlength\bibsep{1pt}

\makeatletter 
\newlength{\figrulesep} 
\setlength{\figrulesep}{0.5\textfloatsep} 

\newcommand{\topfigrule}{\vspace*{-1pt}%
\noindent{\color{cream}\rule[-\figrulesep]{\columnwidth}{1.5pt}} }

\newcommand{\botfigrule}{\vspace*{-2pt}%
\noindent{\color{cream}\rule[\figrulesep]{\columnwidth}{1.5pt}} }

\newcommand{\dblfigrule}{\vspace*{-1pt}%
\noindent{\color{cream}\rule[-\figrulesep]{\textwidth}{1.5pt}} }

\makeatother


\begin{center}
      {\Huge\bfseries
    Supplementary Information
    }
\end{center}
\begin{center}

  \vspace{1.0cm}
  {\LARGE\bfseries
    Hidden excimer formation in the gas-phase photodynamics of a BN-doped phenanthrene\textsuperscript{\dag}
  }

  \vspace{0.5cm}

  {\large Jonas Fackelmayer,\textit{$^{a}$} Michael Bühler,\textit{$^{a}$} Michael Müller,\textit{$^{c}$} Holger Helten,\textit{$^{c}$} Ingo Fischer$^{\ast}$\textit{$^{a}$} and Merle I. S. Röhr$^{\ast}$\textit{$^{a,b}$}}

  \vspace{0.3cm}

  {\small\itshape
    \textsuperscript{a}~Institute of Physical and Theoretical Chemistry,
    University of Würzburg, D-97074 Würzburg, Germany. \\
    
    \textsuperscript{b} Center for Nanosystems Chemistry, University of Würzburg, D-97074 Würzburg, Germany. \\
    
\textsuperscript{c}~Institute of Inorganic Chemistry, University of Würzburg, D-97074 Würzburg, Germany.
  }

  \vspace{0.3cm}

  {\small
    \texttt{ingo.fischer@uni-wuerzburg.de; merle.roehr@uni-wuerzburg.de}
  }

  \vspace{0.2cm}

\end{center}
\thispagestyle{empty}
  

\renewcommand*\rmdefault{bch}\normalfont\upshape
\rmfamily

{
\tableofcontents 
\addtocontents{toc}{\protect\thispagestyle{empty}}
\pagenumbering{arabic}
}
\clearpage



\section{Computational details and optimized geometries} \label{chem_struct}

\begin{figure}[b!]
    \centering
    \includegraphics[width=1\linewidth]{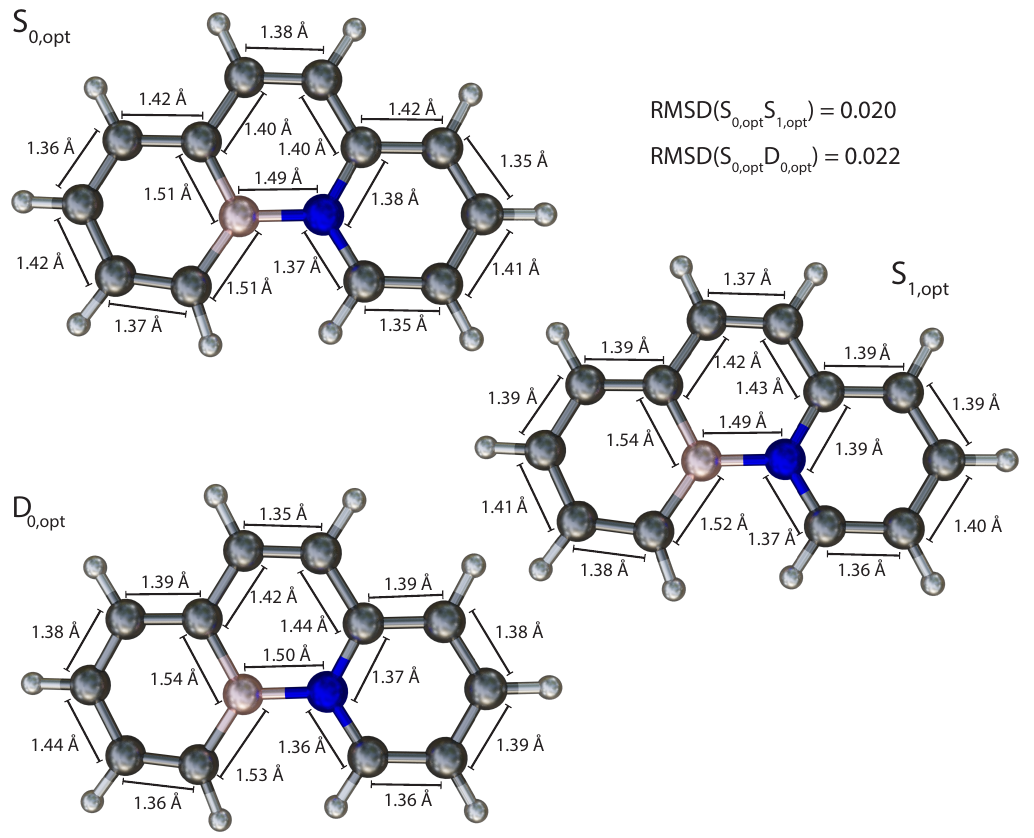}
    \caption{Comparison of optimized molecular geometries in the ground state (S$_{0,\text{opt}}$), first excited state (S$_{1,\text{opt}}$) and cation (D$_{0,\text{opt}}$), computed at the $\omega$B97X-D/aug-cc-pVTZ level of theory. The structural deviations are quantified by the root-mean-square deviation (RMSD) of RMSD(S$_{0\mathrm{,opt}}$S$_{1\mathrm{,opt}}$) = 0.020 Å and RMSD(S$_{0\mathrm{,opt}}$D$_{0\mathrm{,opt}}$) = 0.022 Å.}
    \label{fig:struct_monomer}
\end{figure}

The molecular structure was analyzed within the framework of density functional theory in the gas phase. Optimized geometries for S$_0$, S$_1$ and D$_0$ are shown in Figure~\ref{fig:struct_monomer} and the absence of imaginary vibrational frequencies confirms that these structures correspond to true stationary points on the potential energy surface. The frequency resolved spectrum was computed within the harmonic approximation, explicitly including Herzberg--Teller coupling for the target state, and all simulations were carried out at a temperature of \SI{50}{\kelvin}. The adiabatic Hessian model was employed to define a consistent set of normal coordinates and all related calculations were performed with the \textsc{FCclasses}~3.0 program developed by J.~Cerezo and F.~Santoro.\cite{fabriziosantoroandjaviercerezoFCclasses32019} The resulting spectra were convoluted with a Lorentzian line shape characterized by a half-width at half-maximum of \SI{1}{\milli\electronvolt}. NTOs were obtained from the TD-DFT calculations; cube files were generated with \texttt{cubegen} utility as implemented in \textsc{Gaussian}~16 and plotted with an isovalue of 0.02.\cite{m.j.frischGaussian16Revision2016}

\clearpage

\section{Natural transition orbitals}

\begin{figure}[h!]
    \centering
    \includegraphics[scale=0.9]{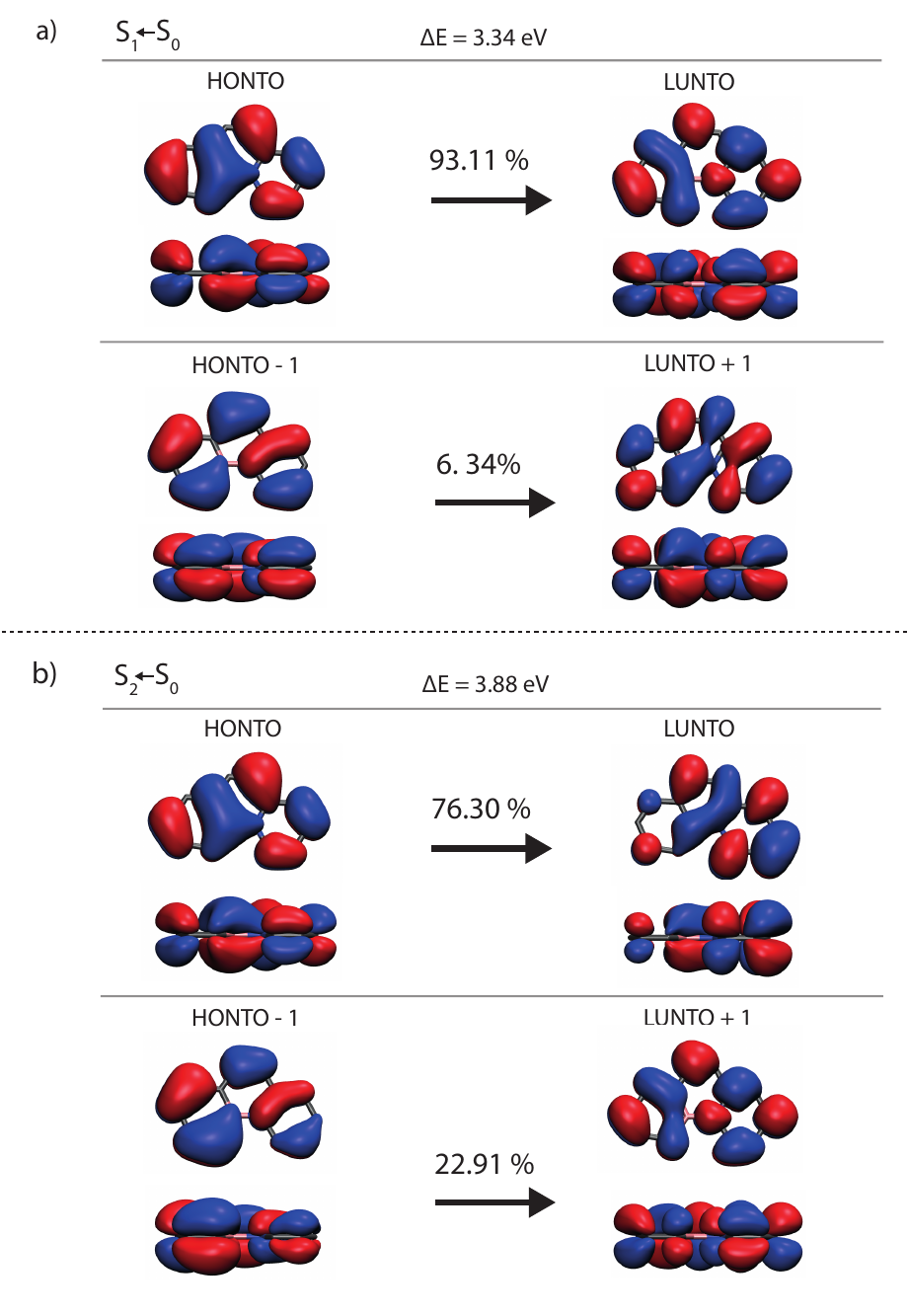}
    \caption{Natural transition orbitals (NTOs) associated with the transitions from the ground state to the first (a) and second (b) excited states are shown. The NTO analysis reveals that both excited states are of predominant $\pi\pi^\ast$ character. All results were obtained from DFT calculations at the $\omega$B97X-D/aug-cc-pVTZ level as implemented in \textsc{Gaussian}~16. The NTOs are visualized using an isovalue of 0.02.}
    \label{fig:NTOs_S1S2}
\end{figure}

\begin{figure}
    \centering
    \includegraphics[]{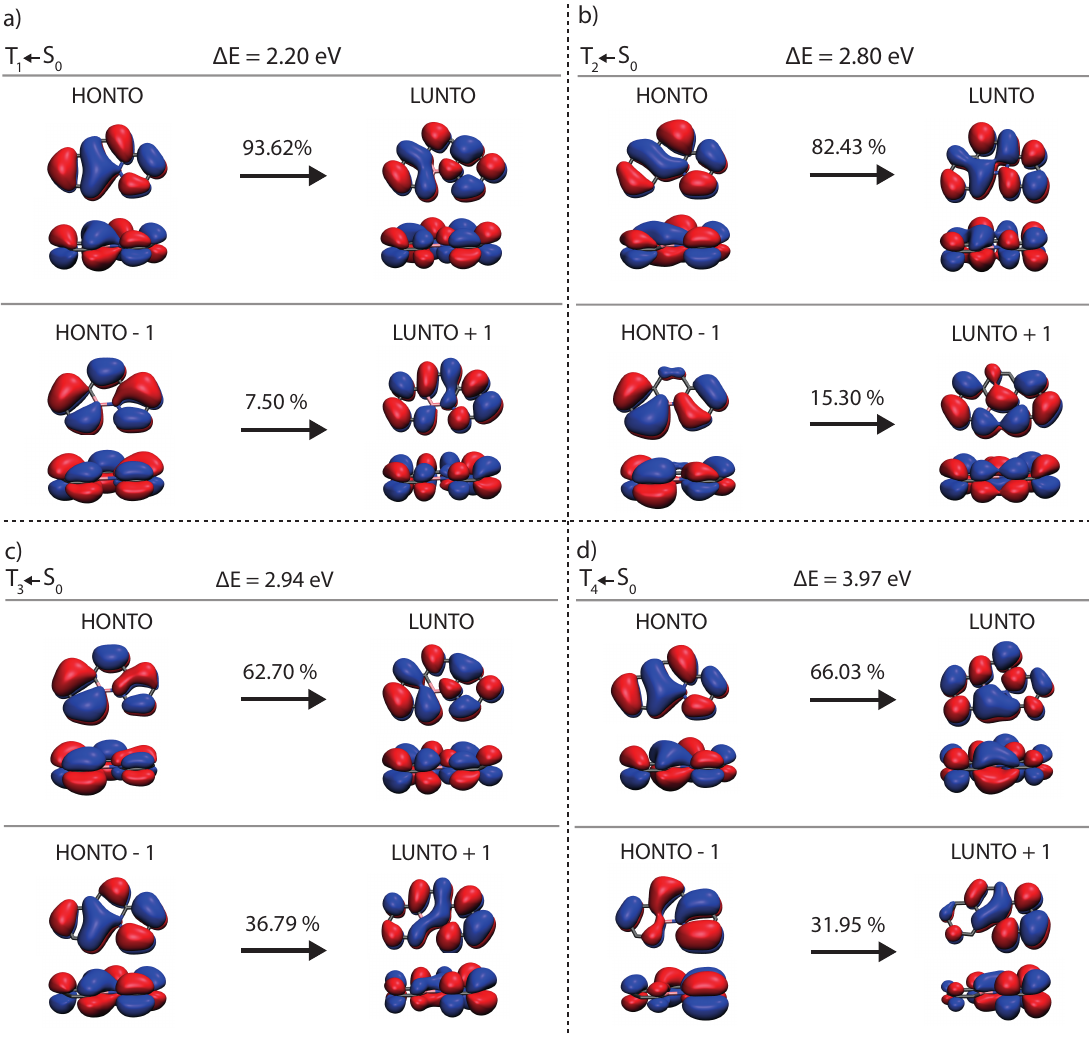}
    \caption{Natural transition orbitals (NTOs) with the main contribution to the transitions between the ground state and the first (a), second (b), third (c), and fourth (d) triplet excited states are shown. The NTO analysis demonstrates that all triplet states exhibit predominant $\pi\pi^\ast$ character. The results are obtained from DFT calculations at the $\omega$B97X-D/aug-cc-pVTZ level using \textsc{Gaussian}~16. The NTOs are visualized with an isovalue of 0.02.}
    \label{fig:NTO_Triplets}
\end{figure}

\begin{figure}
    \centering
    \includegraphics[]{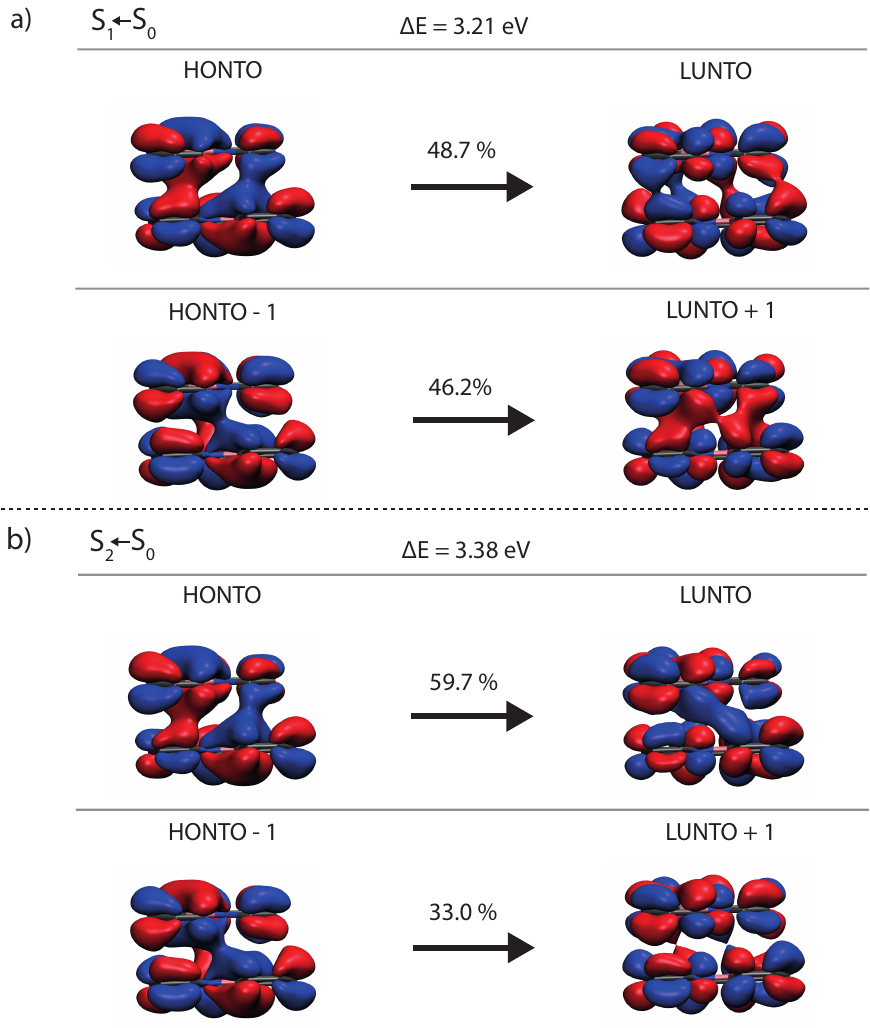}
    \caption{Natural transition orbitals (NTOs) associated with the transitions from the ground state of the optimized minimum-energy dimer structure to the first (a) and second (b) excited states are shown. The NTO analysis reveals that both excited states are of predominant $\pi\pi^\ast$ character, with the orbital overlap clearly visible across the dimer. All results were obtained from DFT calculations at the $\omega$B97X-D3/aug-cc-pVDZ level using ORCA~6.0.1.\cite{neeseSoftwareUpdateORCA2025} The NTOs are visualized using an isovalue of 0.02.}
    \label{fig:DIMER_NTOs}
\end{figure}
\clearpage
\section{Frequencies of S$_0$, S$_1$ and D$_0$}

\begin{table}[ht]
  \centering  \captionsetup{justification=centering,singlelinecheck=false}
  \caption{Harmonic vibrational frequencies of the S$_0$, S$_1$, and D$_0$ monomer states (in cm$^{-1}$) of \textbf{1} ($\omega$B97X-D/aug-cc-pVTZ).}
  \begin{tabular}{cccc@{\hspace{1.5em}}cccc}
    \toprule
    Vibrational mode & \multicolumn{3}{c}{Frequency / cm$^{-1}$} &
    Vibrational mode & \multicolumn{3}{c}{Frequency / cm$^{-1}$} \\
    \cmidrule(lr){2-4}\cmidrule(lr){6-8}
    & S$_0$ & S$_1$ & D$_0$ & & S$_0$ & S$_1$ & D$_0$ \\
    \midrule
    $\nu_{66}$ &   95.48 &   75.28 &   80.80 & $\nu_{33}$ & 1051.27 & 1004.01 & 1079.68 \\
    $\nu_{65}$ &  100.71 &   81.59 &   90.59 & $\nu_{32}$ & 1102.10 & 1069.98 & 1106.54 \\
    $\nu_{64}$ &  210.32 &  182.01 &  201.74 & $\nu_{31}$ & 1133.12 & 1110.55 & 1123.96 \\
    $\nu_{63}$ &  231.40 &  191.91 &  211.00 & $\nu_{30}$ & 1180.04 & 1124.26 & 1179.26 \\
    $\nu_{62}$ &  232.08 &  201.33 &  231.60 & $\nu_{29}$ & 1185.12 & 1162.52 & 1182.80 \\
    $\nu_{61}$ &  397.08 &  325.79 &  364.11 & $\nu_{28}$ & 1195.67 & 1168.56 & 1198.06 \\
    $\nu_{60}$ &  402.23 &  384.79 &  388.02 & $\nu_{27}$ & 1221.46 & 1182.57 & 1209.75 \\
    $\nu_{59}$ &  418.16 &  391.43 &  401.20 & $\nu_{26}$ & 1223.45 & 1205.79 & 1228.79 \\
    $\nu_{58}$ &  419.01 &  405.86 &  405.00 & $\nu_{25}$ & 1267.59 & 1252.77 & 1265.52 \\
    $\nu_{57}$ &  474.49 &  426.77 &  446.30 & $\nu_{24}$ & 1295.04 & 1256.98 & 1290.14 \\
    $\nu_{56}$ &  492.38 &  466.56 &  466.34 & $\nu_{23}$ & 1309.70 & 1280.42 & 1305.94 \\
    $\nu_{55}$ &  520.22 &  488.99 &  481.51 & $\nu_{22}$ & 1388.67 & 1362.02 & 1378.61 \\
    $\nu_{54}$ &  546.62 &  520.99 &  532.96 & $\nu_{21}$ & 1399.14 & 1393.17 & 1414.09 \\
    $\nu_{53}$ &  592.85 &  554.77 &  557.95 & $\nu_{20}$ & 1421.78 & 1407.92 & 1444.92 \\
    $\nu_{52}$ &  608.29 &  594.50 &  597.27 & $\nu_{19}$ & 1448.30 & 1435.17 & 1471.17 \\
    $\nu_{51}$ &  697.40 &  667.33 &  682.35 & $\nu_{18}$ & 1466.14 & 1460.74 & 1478.79 \\
    $\nu_{50}$ &  712.60 &  679.19 &  693.51 & $\nu_{17}$ & 1499.24 & 1489.75 & 1495.56 \\
    $\nu_{49}$ &  731.57 &  701.99 &  728.85 & $\nu_{16}$ & 1534.87 & 1523.71 & 1533.56 \\
    $\nu_{48}$ &  746.26 &  726.64 &  746.15 & $\nu_{15}$ & 1563.05 & 1535.54 & 1541.34 \\
    $\nu_{47}$ &  780.34 &  729.69 &  783.27 & $\nu_{14}$ & 1597.52 & 1586.39 & 1602.58 \\
    $\nu_{46}$ &  794.82 &  741.87 &  806.44 & $\nu_{13}$ & 1617.92 & 1608.74 & 1617.55 \\
    $\nu_{45}$ &  836.12 &  764.95 &  824.16 & $\nu_{12}$ & 1658.45 & 1594.32 & 1643.80 \\
    $\nu_{44}$ &  837.75 &  794.25 &  850.08 & $\nu_{11}$ & 1714.25 & 1660.33 & 1688.72 \\
    $\nu_{43}$ &  864.67 &  820.68 &  877.41 & $\nu_{10}$ & 3143.85 & 3147.69 & 3184.73 \\
    $\nu_{42}$ &  887.84 &  826.09 &  892.97 & $\nu_{9}$ & 3149.22 & 3156.97 & 3189.60 \\
    $\nu_{41}$ &  895.02 &  854.63 &  933.84 & $\nu_{8}$ & 3150.90 & 3168.45 & 3192.58 \\
    $\nu_{40}$ &  987.78 &  855.82 &  988.91 & $\nu_{7}$ & 3165.65 & 3171.37 & 3201.44 \\
    $\nu_{39}$ &  988.88 &  939.51 &  998.10 & $\nu_{6}$ & 3198.59 & 3199.68 & 3217.22 \\
    $\nu_{38}$ & 1001.10 &  939.78 & 1004.20 & $\nu_{5}$ & 3206.98 & 3206.99 & 3222.32 \\
    $\nu_{37}$ & 1016.97 &  952.61 & 1017.93 & $\nu_{4}$ & 3214.54 & 3211.13 & 3224.98 \\
    $\nu_{36}$ & 1022.77 &  969.42 & 1031.27 & $\nu_{3}$ & 3224.22 & 3219.79 & 3236.87 \\
    $\nu_{35}$ & 1024.13 &  974.44 & 1051.86 & $\nu_{2}$ & 3234.23 & 3233.51 & 3245.52 \\
    $\nu_{34}$ & 1033.32 &  987.57 & 1057.82 & $\nu_{1}$ & 3262.96 & 3256.15 & 3269.80 \\
    \bottomrule
  \end{tabular}
  \label{tab:freq_S0_S1_D0}
\end{table}

\clearpage

\section{Assignments for the vibrational bands of the REMPI spectrum}

\begin{table}[h]
\centering
\caption{Summary of selected bands of the S$_1$ state of \textbf{1}. Tentative mode assignments are based on DFT ($\omega$B97X-D/aug-cc-pVTZ) calculations (see Tab.~\ref{tab:freq_S0_S1_D0}). The calculated zero-point-energy-corrected S$_1$ origin at 24964~cm$^{-1}$ is blue-shifted by 2084~cm$^{-1}$ relative to the experimental origin at 22880~cm$^{-1}$. All other calculated energies are given relative to this origin. The assignments are based on a comparison of the experimental spectrum with the shifted line spectrum shown in Fig.~\ref{fig:VibExpVCalc}.}
\begin{tabular}{lll}
\toprule
Vib. Energy / cm$^{-1}$ & Calc. Energy / cm$^{-1}$ & Tentative assignment  \\
\midrule
22880	& 24964          	& S$_1$ origin \\
+135  	& +151 and +162   	& $\tilde{\nu}^2_{66}$, $\tilde{\nu}^2_{65}$ out-of-plane ring deform. \\
+220  	& +191        		& $\tilde{\nu}_{63}$ in-plane ring deform.\\
+343  	& +364        		& $\tilde{\nu}^2_{64}$ out-of-plane ring deform.\\
+381  	& +401        		& $\tilde{\nu}_{66}$ + $\tilde{\nu}_{61}$ out-of-plane ring deform. comb.\\
+452  	& +467        		& $\tilde{\nu}_{56}$ in-plane ring deform. \\
+506  	& +521        		& $\tilde{\nu}_{54}$ in-plane ring deform. \\
+572  	& +594        		& $\tilde{\nu}_{52}$ in-plane ring deform. \\
+655  	& +679        		& $\tilde{\nu}_{50}$ BN/BC-stretch + in-plane ring deform. \\
+705  	& +730        		& $\tilde{\nu}_{47}$ in-plane ring deform.\\
+794  	& --         		& -- \\
+850  	& +940        		& $\tilde{\nu}_{39}$ C–H-out-of-plane deform.\\
+975  	& --         		& -- \\
+1066 	& --         		& -- \\
+1153 	& --         		& -- \\
+1187 	& +1168       		& $\tilde{\nu}_{28}$ C–H-in-plane deform.\\
+1333 	& +1395 and +1422 	& $\tilde{\nu}_{20}$, $\tilde{\nu}_{18}$ BC-stretch + C–H-in-plane deform. \\
+1457 	& --         		& -- \\
+1492 	& --         		& -- \\
+1552 	& --         		& -- \\
+1787 	& --         		& -- \\
+1848 	& --         		& -- \\
+2014 	& --         		& -- \\
+2070 	& --         		& -- \\
\bottomrule
\end{tabular}

\label{tab:VibExpVCalc}
\end{table}

\begin{figure}[h!]
    \centering
    \includegraphics[width=14cm]{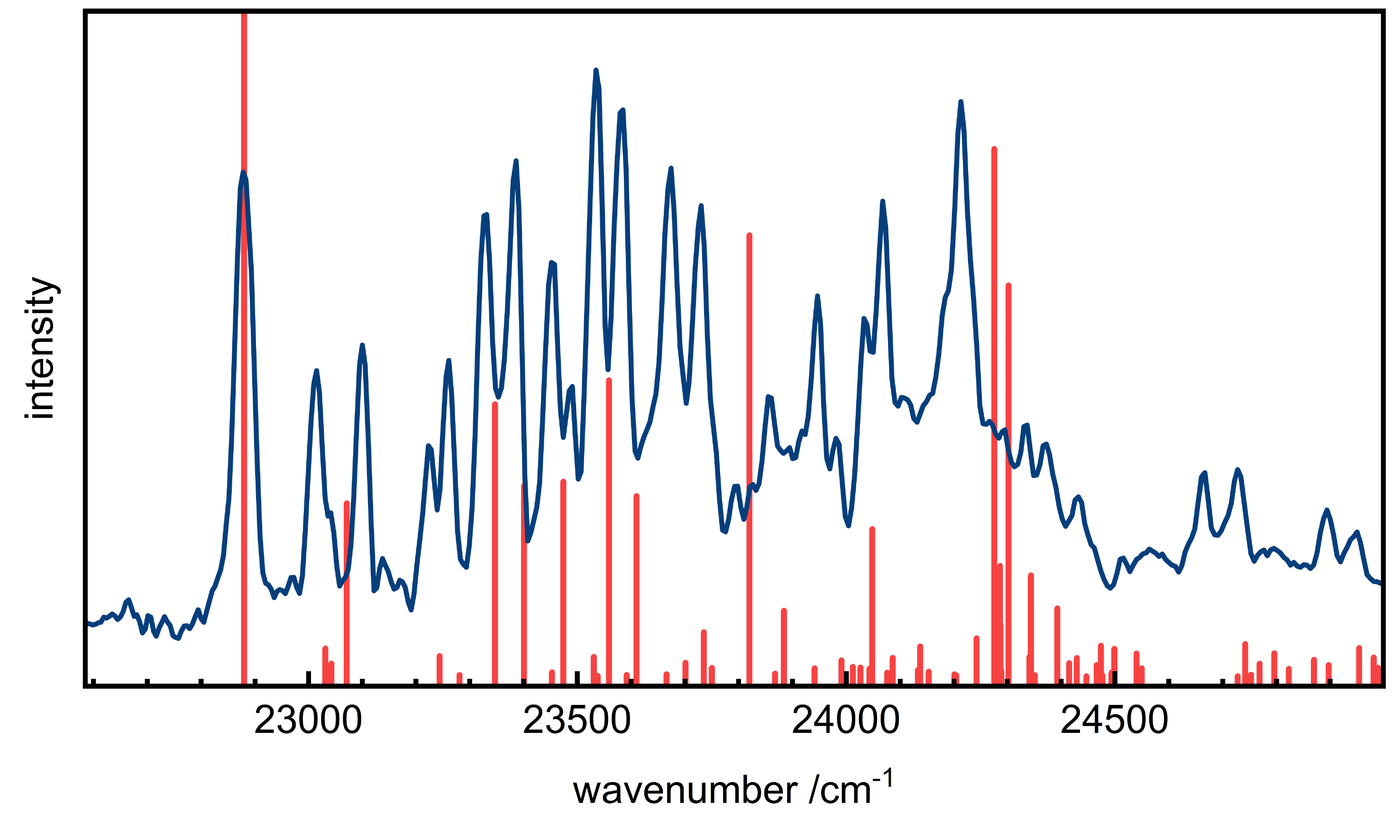}
    \caption{Experimental [1+1$^\prime$] REMPI spectrum of \textbf{1} in blue with the calculated line spectrum in red. The line spectrum was shifted by 2084 cm$^{-1}$ to match the 0\textsuperscript{0} transition. The stick spectrum intensities were magnified to better visualize the bands assigned in Tab.\ref{tab:VibExpVCalc}.}
    \label{fig:VibExpVCalc}
\end{figure}

\clearpage
\section{Spin-orbit coupling and intersystem crossing rates}


Spin-orbit coupling (SOC) matrix elements were computed between the electronic ground state and the first two excited singlet states (S$_0$, S$_1$, S$_2$) and the energetically relevant triplet states (T$_1$-T$_5$). All calculations were performed with the range-separated hybrid functional $\omega$B97X-D3 and the aug-cc-pVTZ basis set as implemented in ORCA~6.0.1.\cite{neeseSoftwareUpdateORCA2025} For the optimized S$_0$ geometry, SOC values were evaluated for all S$_j$/T$_n$ pairs (see Tab.~\ref{tab:SOC_opt}).\cite{neeseEfficientAccurateApproximations2005,desouzaPredictingPhosphorescenceRates2019} At this geometry, the largest SOC amounts to only \SI{0.13}{\per\centi\meter} and is found between S$_0$ and T$_1$, while all other couplings are even smaller, indicating intrinsically weak singlet-triplet mixing at the Franck-Condon region.

\begin{figure}
    \centering
    \includegraphics[width=1.0\linewidth]{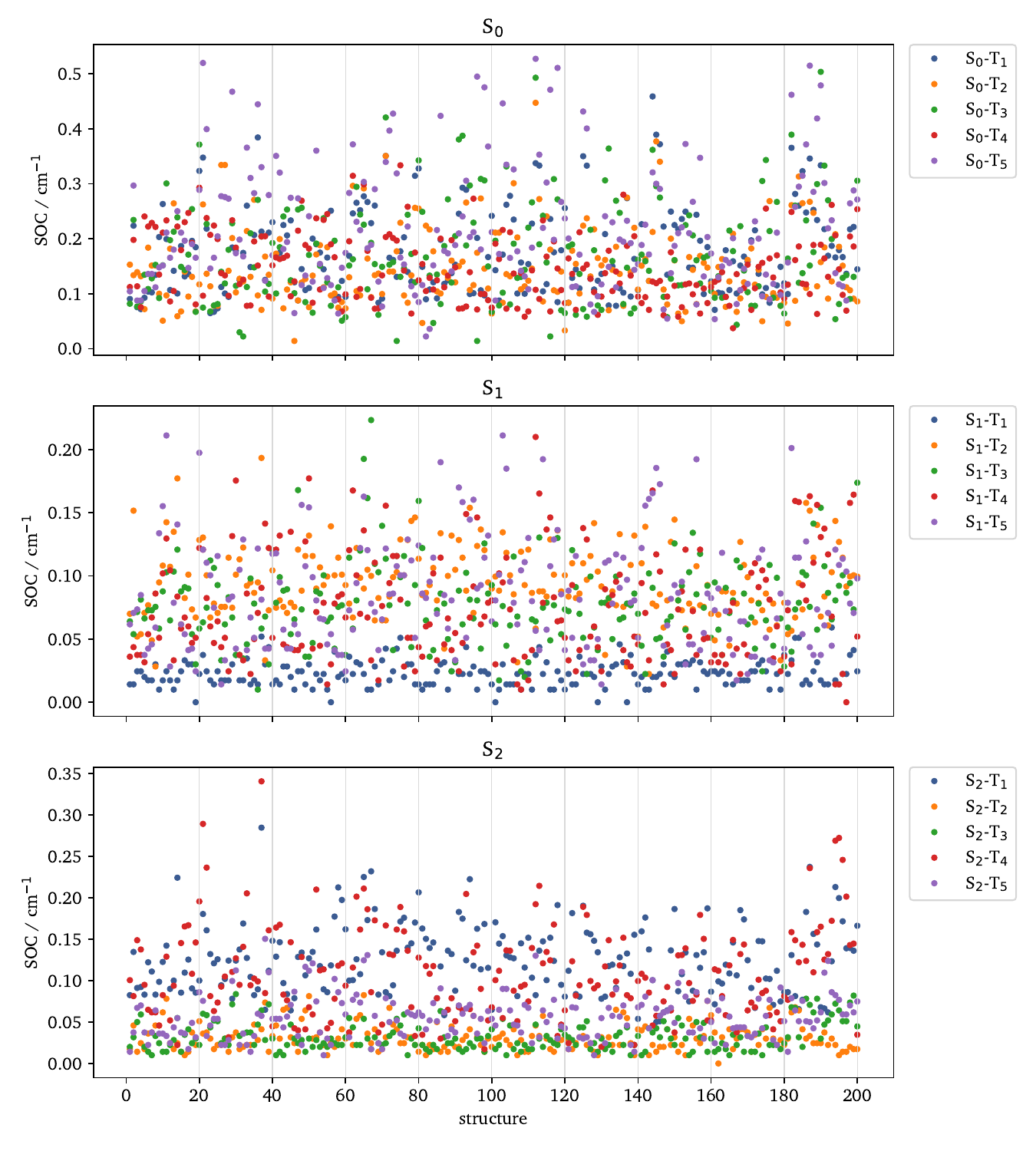}
    \caption{Spin-orbit coupling matrix elements between S$_0$, S$_1$, and S$_2$ and the low-lying triplet states T$_1$--T$_5$ along the molecular dynamics trajectories. The top, middle, and bottom panels correspond to S$_0$, S$_1$, and S$_2$, respectively, and show the SOC values (in \si{\per\centi\meter}) computed at the $\omega$B97X-D3/aug-cc-pVTZ level for individual MD snapshots.}
    \label{fig:SOC}
\end{figure}

To assess whether thermally induced structural distortions could significantly enhance the SOC, a ground-state molecular dynamics simulation (MD) was carried out. An xTB-based MD on the S$_0$ surface was run at \SI{50}{\kelvin} for \SI{1}{\pico\second} with a \SI{0.5}{\femto\second} time step using a Nosé--Hoover chain thermostat at \SI{50}{\kelvin}. From this trajectory, 12 representative trajectories were selected and subsequently propagated for an additional \SI{500}{\femto\second} at the $\omega$B97X-D3/aug-cc-pVTZ level.\cite{weigendBalancedBasisSets2005,linLongRangeCorrectedHybrid2013} Along these DFT-based trajectories, TD-DFT calculations for singlet and triplet states including SOC were performed for 20 randomly selected snapshots per trajectory after \SI{200}{\femto\second} (see Fig.~\ref{fig:SOC}). Even under these distorted geometries, the maximum SOC value observed is only about \SI{0.50}{\per\centi\meter}, confirming that structurally accessible configurations do not exhibit substantially stronger SOC.

In a further step, attempts were made to optimize both the relevant singlet and triplet excited states in order to obtain intersystem crossing (ISC) rate constants.\cite{desouzaPredictingPhosphorescenceRates2019} Starting from selected structures along the trajectories, excited-state optimizations with state tracking were performed using the ORCA keywords \texttt{FOLLOWIROOT true} and \texttt{FIRKEEPFIRSTREF true}. This yielded converged geometries and corresponding ISC rates for the S$_1$--T$_1$ and S$_1$--T$_2$ pairs. For the energetically closest triplet state T$_3$, however, no stable optimized structure could be obtained: with state tracking, the optimization oscillates between T$_3$ and T$_2$, while optimizations without state tracking consistently evolve toward a state with T$_2$ character. Consequently, an ISC rate for S$_1\rightarrow$T$_3$ could not be computed. Given the consistently weak SOC values found for T$_3$ along the MD snapshots (see Fig.~\ref{fig:SOC}), it is nevertheless reasonable to assume that the ISC rate to this state is unlikely to exceed those obtained for the other triplet states.
\begin{table}[ht]
\captionsetup{justification=centering,singlelinecheck=false}
  \centering
  \caption{Spin-orbit coupling matrix elements between S$_j$ and T$_n$ (in cm$^{-1}$).}
  \begin{tabular}{cccc}
    \toprule
    Electronic state & \multicolumn{3}{c}{SOCME / cm$^{-1}$} \\
    \cmidrule(lr){2-4}
    T$_n$ & S$_0$ & S$_1$ & S$_2$ \\
    \midrule
    T$_1$ & 0.13 & 0.01 & 0.06 \\
    T$_2$ & 0.09 & 0.07 & 0.02 \\
    T$_3$ & 0.04 & 0.03 & 0.01 \\
    T$_4$ & 0.08 & 0.01 & 0.02 \\
    T$_5$ & 0.04 & 0.03 & 0.01 \\
    \bottomrule
  \end{tabular}
  \label{tab:SOC_opt}
\end{table}

\begin{table}[ht]
\captionsetup{justification=centering,singlelinecheck=false}
  \centering
  \caption{Intersystem crossing rates, given in \si{\per\second}, between S$_1$ and the relevant triplet states. The median, mean, minimum and maximum rates obtained from the sampled geometries are listed, whereas the individual values are shown in Fig.~\ref{fig:ISC}.}
  \begin{tabular}{ccccc}
    \toprule
    Target    & Median    & Mean      & Min       & Max       \\
    \midrule
    T$_1$ & 1.57e+04  & 3.62e+04  & 6.53e+01  & 1.46e+05  \\
    T$_2$ & 9.58e+05  & 1.15e+06  & 1.65e+05  & 2.63e+06  \\
    \bottomrule
  \end{tabular}
  \label{tab:ISC_T}
\end{table}
 
\begin{figure}
    \centering
    \includegraphics[width=1.0\linewidth]{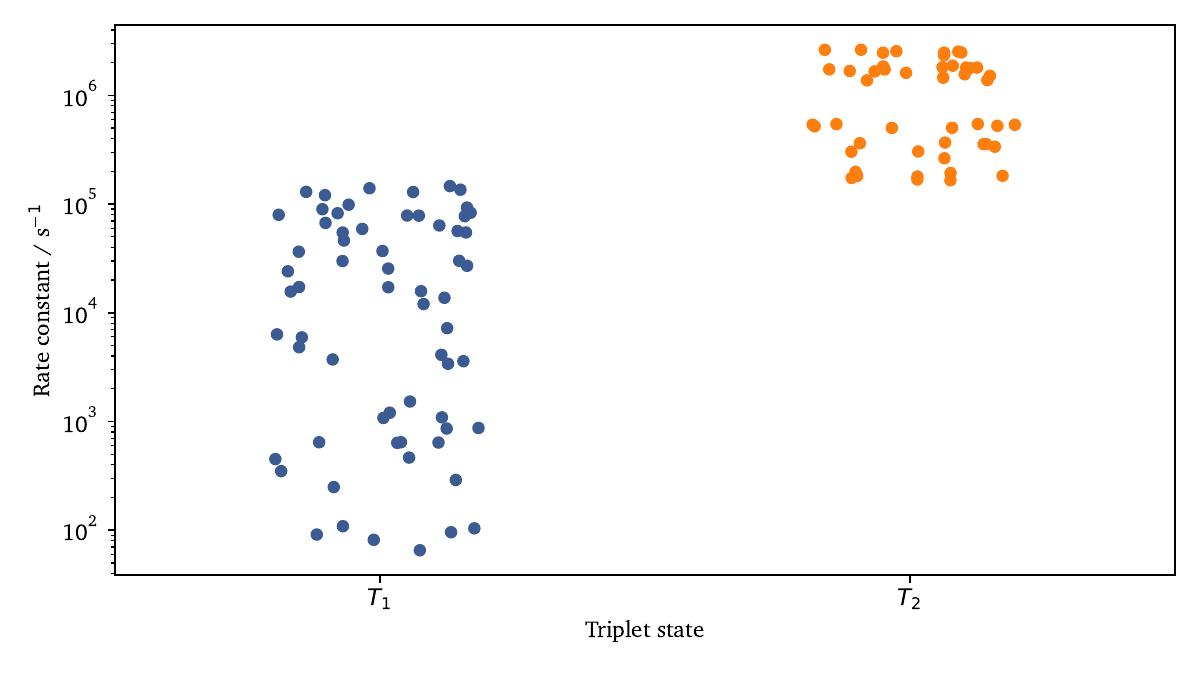}
    \caption{Intersystem crossing rate constants from S$_1$ to the triplet states T$_1$ and T$_2$. Each point corresponds to a calculated rate constant for an individual structure, plotted on a logarithmic scale. Rates were obtained at the $\omega$B97X-D3/aug-cc-pVTZ level and are given in \si{\per\second}.}
    \label{fig:ISC}
\end{figure}

 \clearpage

\section{Comparison of liquid-phase absorption with spectral simulations}

\begin{figure}[h]
    \centering
    \includegraphics[]{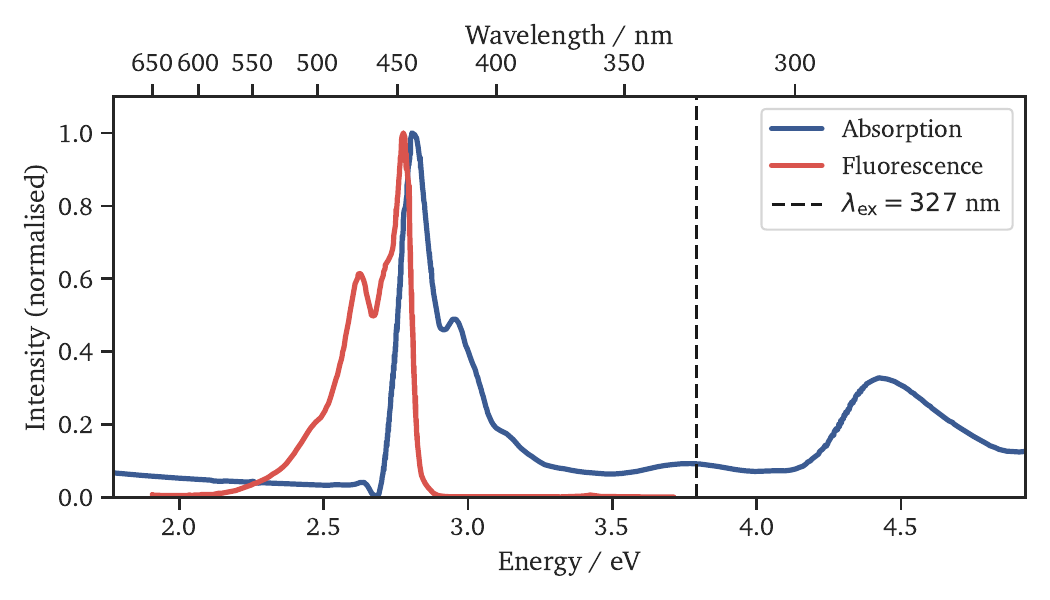}
    \caption{Normalized absorption (blue) and fluorescence emission (red) spectra of the molecule in THF, plotted as a function of photon energy, with the corresponding wavelength scale shown on the upper axis. The fluorescence spectrum was recorded at an excitation wavelength of 327~nm, indicated by the dashed line. The shoulder of the absorption band can be attributed to a vibrational progression in the S$_1$ state (see Figs.~\ref{fig:calc_vs_exp} and \ref{fig:stick_broadened}). Adapted with permission from Ref.\citenum{mullerBNPhenanthreneBNPyreneBasedFluorescent2024}.}
    \label{fig:abs_fluor}
\end{figure}

\begin{figure}[h!]
    \centering
    \includegraphics[]{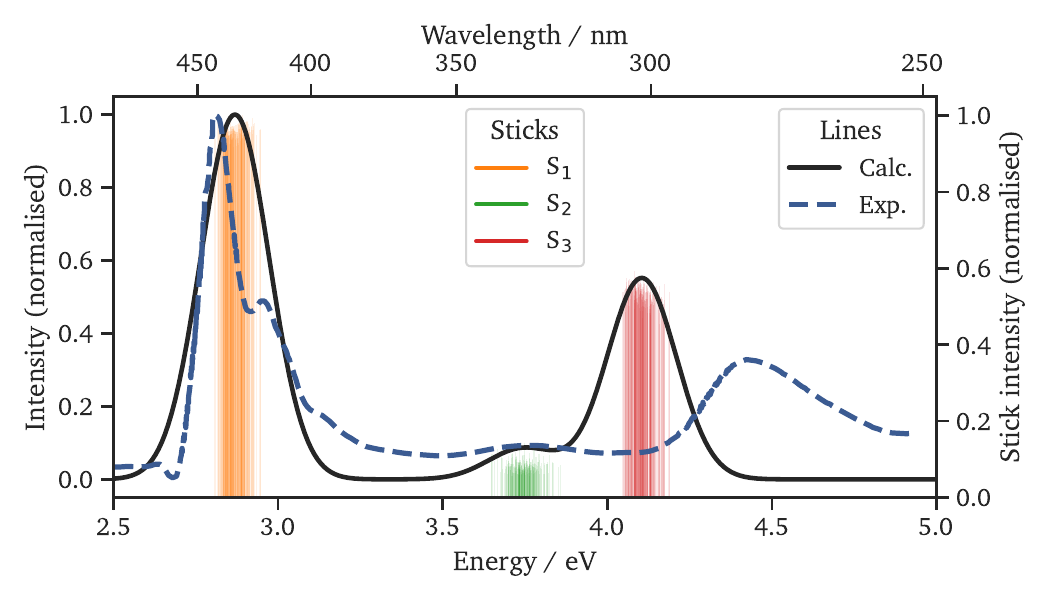}
    \caption{Calculated ensemble absorption spectrum of BN-phenanthrene (black), computed using the conductor-like polarizable continuum model for THF,\cite{baroneQuantumCalculationMolecular1998,garcia-ratesEffectSoluteCavity2020} compared with the experimental spectrum measured in THF (blue), shown as normalized intensity versus energy (bottom axis) and wavelength (top axis). The stick spectrum (top) highlights the contributions of the lowest excited states, revealing a clear energetic splitting between the S$_1$ (orange), S$_2$ (green), and S$_3$ (red) manifolds.}
    \label{fig:calc_vs_exp}
\end{figure}

\begin{figure}
    \centering
    \includegraphics[width=1.0\linewidth]{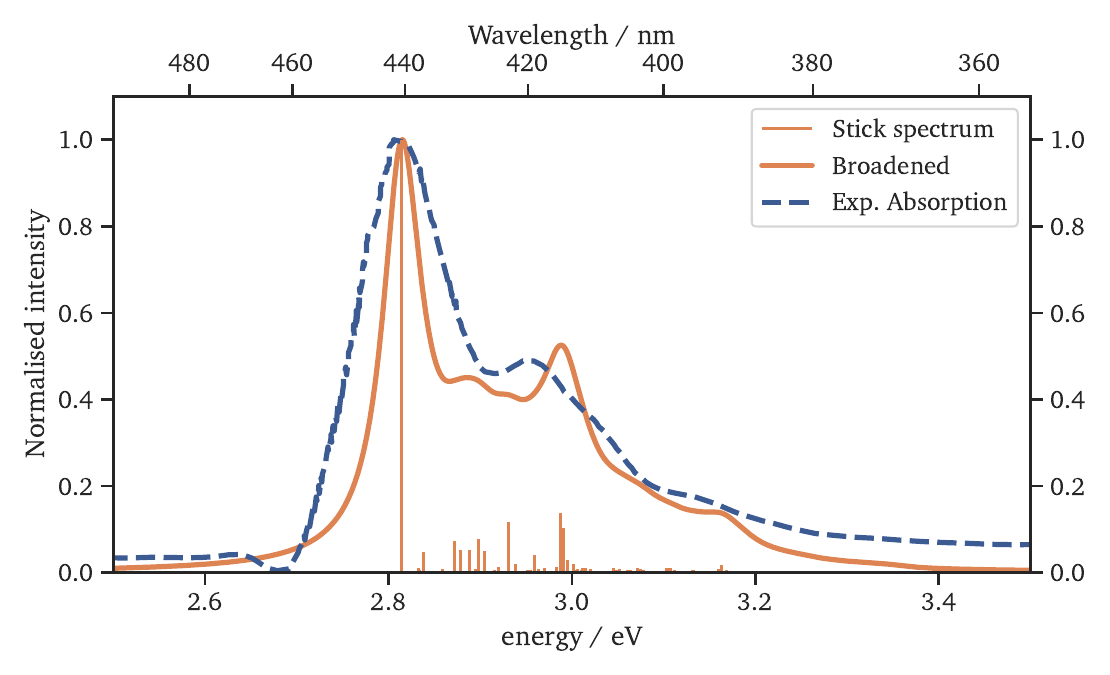}
    \caption{Frequency resolved computed absorption spectrum shown as a stick spectrum and a broadened profile, overlaid with the experimental absorption spectrum measured in THF. An enlarged view of the absorption band shown in Fig.~\ref{fig:calc_vs_exp} is displayed to resolve individual vibronic features. The comparison shows that the band maximum at 2.67\,eV (465\,nm) and the two subsequent shoulders originate from a vibrational progression of the S$_1$ state.}
    \label{fig:stick_broadened}
\end{figure}

\clearpage
\section{Time-resolved mass spectra} \label{sec:ExData}

\begin{figure}[h!]
    \centering
    \includegraphics[width=10cm]{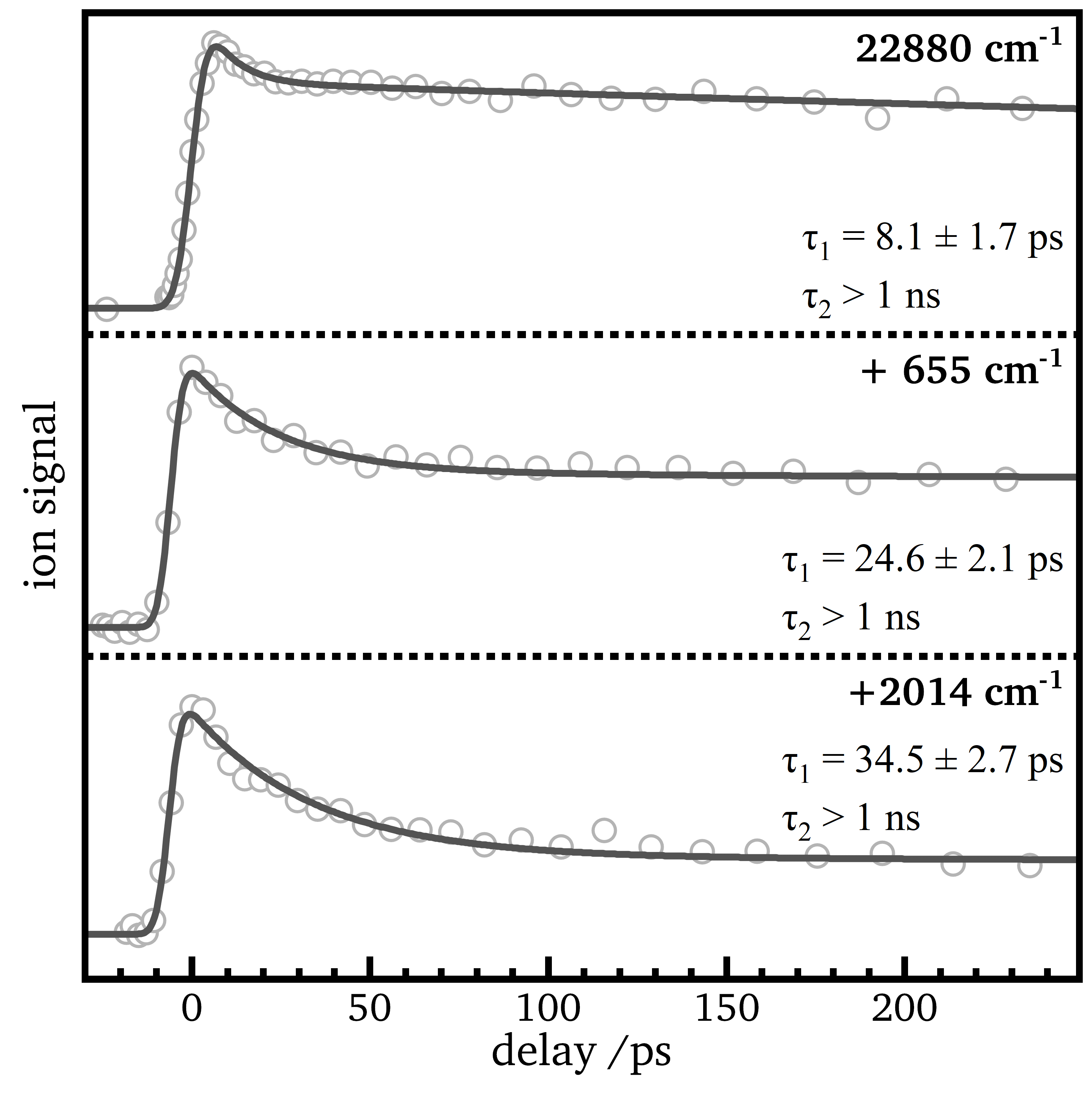}
    \caption{Delay traces of the ion signal at the S$_1$ band origin (top trace) and at two different excitation energies. A ps-time dependence is visible at least at the two higher excitation energies.}
    \label{fig:IonTraces}
\end{figure}

\begin{figure}[h!]
    \centering
    \includegraphics[width=12cm]{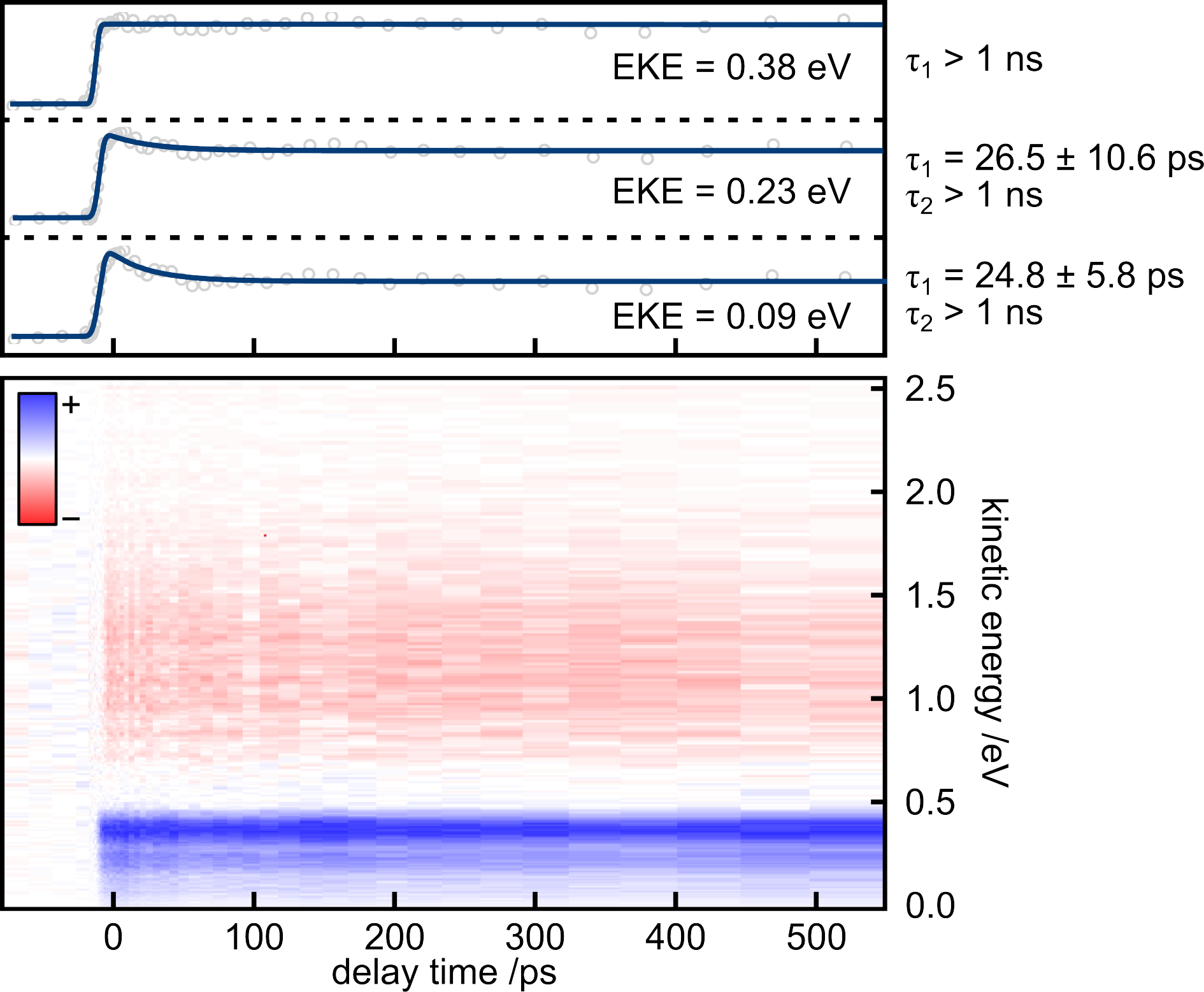}
    \caption{Time-resolved map of the photoelectron signal, recorded upon excitation with 2.90 eV (427.7 nm). Time-independent background signals from pump and probe laser have been subtracted. Blue bands signify an increase of signal after interaction with the pump laser while red bands are a result of signal decrease. At higher eKE above 0.5 eV, weak signals originating from two-photon excitation are visible. Top: Delay traces of electrons at specified kinetic energies.}
    \label{fig:TRPES_427.7nm}
\end{figure}

\begin{figure}[h!]
    \centering
    \includegraphics[width=10cm]{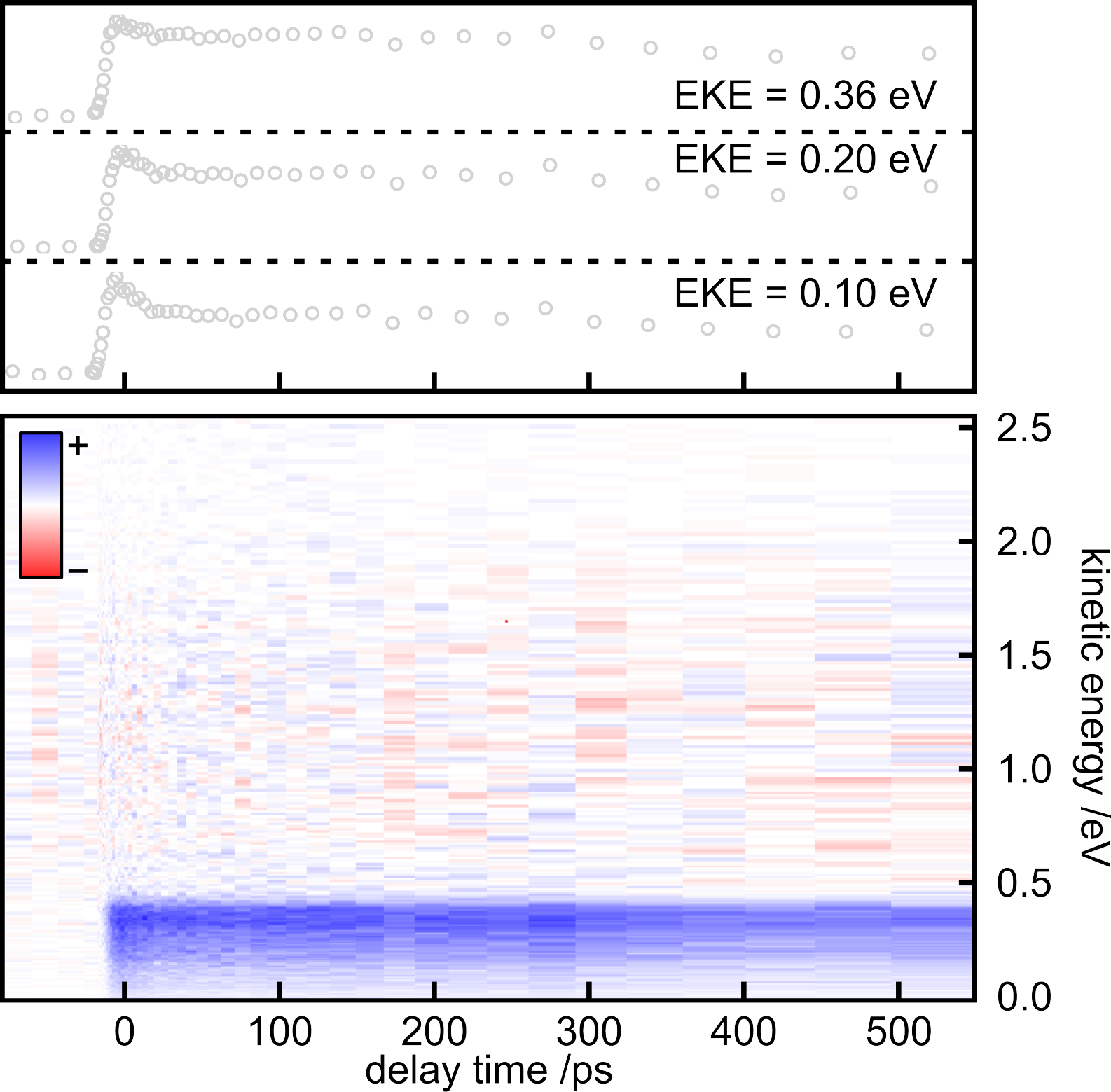}
    \caption{Time-resolved map of the photoelectron signal, recorded upon excitation with 3.09 eV (401.6 nm). Time-independent background signals from pump and probe laser have been subtracted. Blue bands signify an increase of signal after interaction with the pump laser while red bands are a result of signal decrease. Top: Delay data of electrons at specified kinetic energies.}
    \label{fig:TRPES_401.6nm}
\end{figure}

In addition to the time-resolved photoelectron measurement, recorded at the S$_1$ origin, similar experiments were conducted at higher excitation energies (see Fig. \ref{fig:TRPES_427.7nm} and \ref{fig:TRPES_401.6nm}). As observed for the  0$^0$ transition, the signal shows three distinct bands at different kinetic energies. For an excitation energy of 2.90 eV the most intense band at 0.38 eV shows a time constant on the ns-scale while the bands of lower energy match the time constant of the ion signal decay (see Fig. \ref{fig:IonTraces}). The data recorded at an excitation energy of 3.09 eV exhibit a low signal-to-noise ratio and no reliable fits were obtained.

\clearpage

\section{4a,4b-Azaboraphenanthrene dimer}

\subsection{TOF spectrum}
\begin{figure}[h]
    \centering
    \includegraphics[width=12cm]{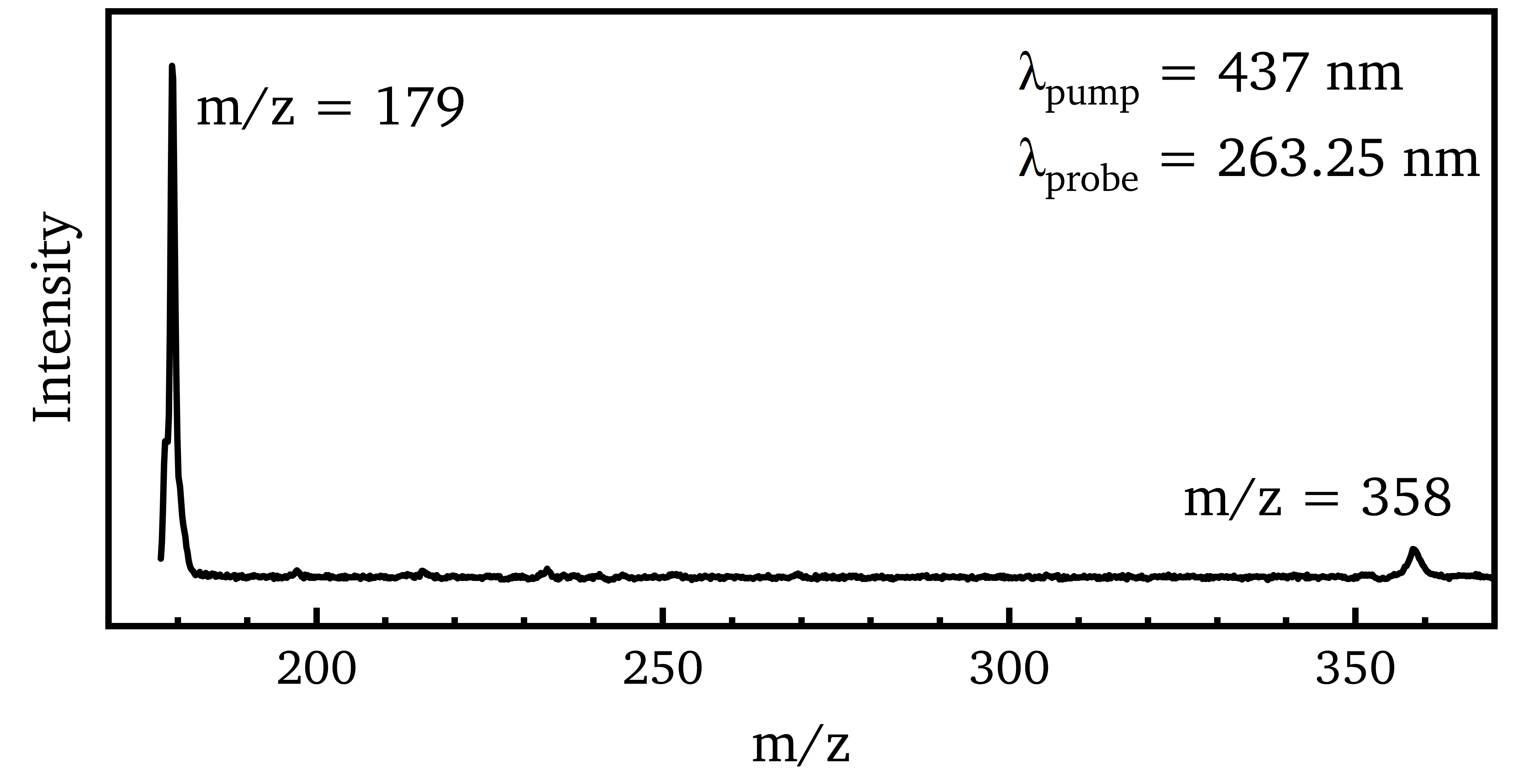}
    \caption{TOF spectrum under conditions optimized for dimer formation. The changes in the conditions consisted of an increase in heating of the sample (128 to 135$^{\circ}$C), an increase of the pulse width (96 to 126 $\mu$s) and a decrease in backing pressure (1.2 to 1.0 bar). }
    \label{fig:Tof_Dimer}
\end{figure}

\subsection{Wavelength dependence of the TER distribution}
\begin{figure}[h]
    \centering
    \includegraphics[width=12cm]{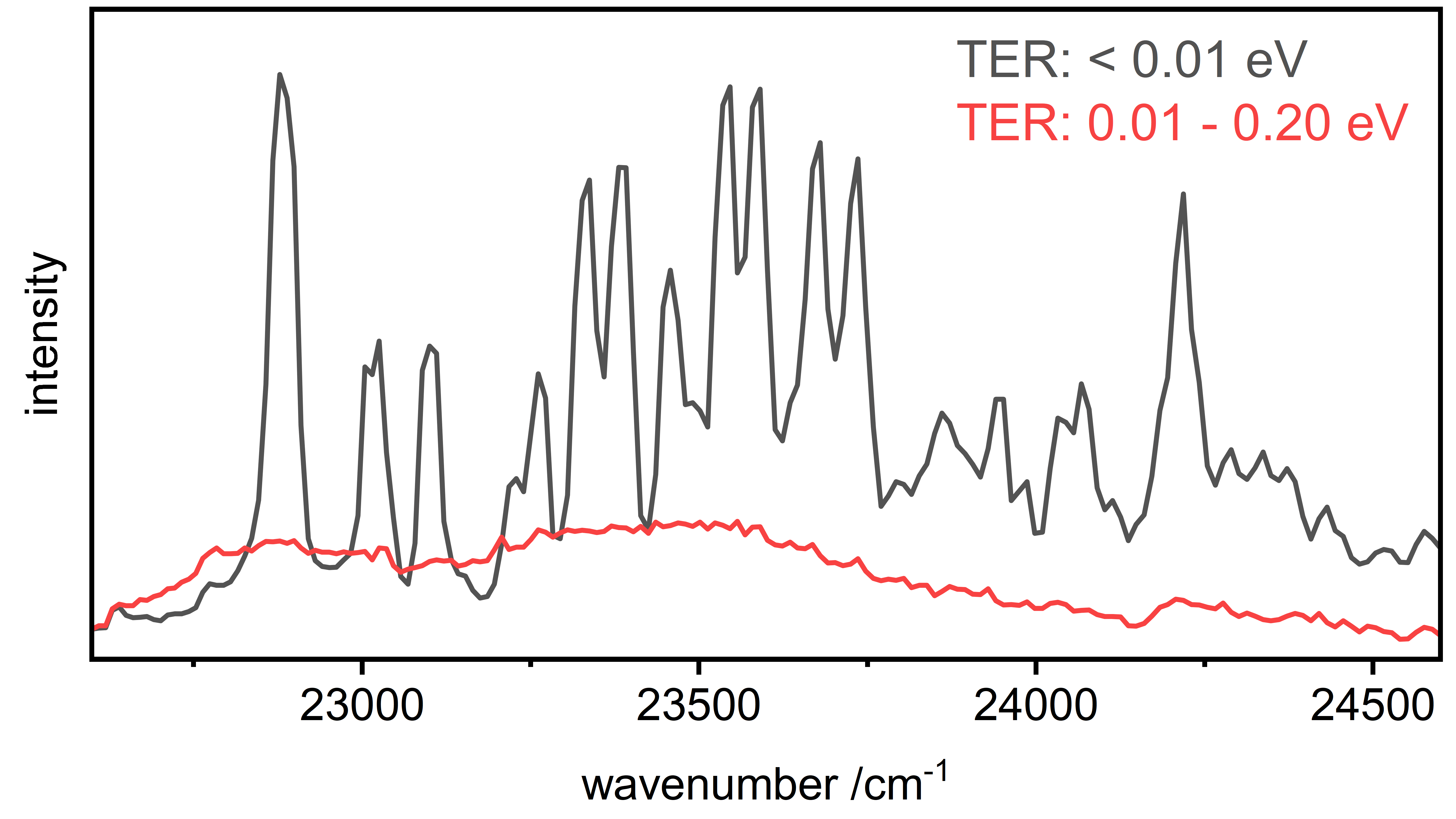}
    \caption{Difference in signal intensity of the ions (m/z = 178/179) as depending on the TER distribution. The black trace shows the behavior of the ions at close to zero TER with sharp, well-defined peaks. The red trace shows a broad unstructured signal that can be attributed to preceding fragmentation of the ion.}
    \label{fig:REMPI_Dimer}
\end{figure}
\clearpage
\subsection{Dimer search}

\begin{figure}[h]
    \centering
    \includegraphics[]{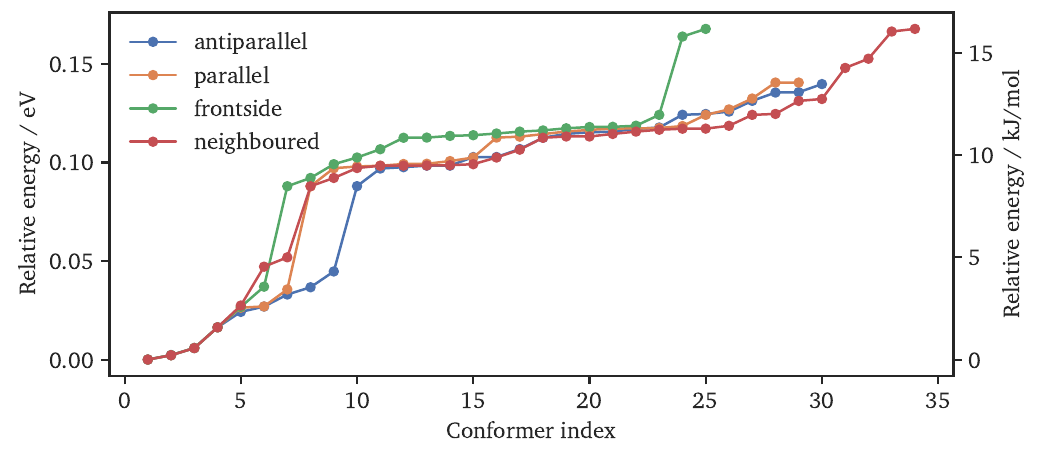}
    \caption{Relative energies of the conformers obtained from the CREST search using \texttt{xTB}. The structures are grouped according to the initial dimer arrangement used in the search, namely antiparallel, parallel, frontside and laterally displaced configurations. Energies are given relative to the lowest energy conformer.}
    \label{fig:CREST}
\end{figure}

The conformational search was performed using CREST 3.0\cite{prachtCRESTProgramExploration2024} with GFN2-xTB,\cite{bannwarthGFN2xTBAnAccurateBroadly2019} as implemented in the xTB program,\cite{bannwarthExtendedTightbindingQuantum2021} employing four distinct initial dimer arrangements. These included H-aggregate-type structures with parallel alignment of the $\pi$-cores, H-aggregates with antiparallel orientation, dimers aligned according to their dipole moments in a face-to-face arrangement, and laterally displaced neighboring dimers. The structures, as shown in Fig.~\ref{fig:CREST}, obtained from the CREST search, 118 in total, were subsequently superimposed to remove duplicates.

\begin{figure}[h!]
    \centering
    \includegraphics[]{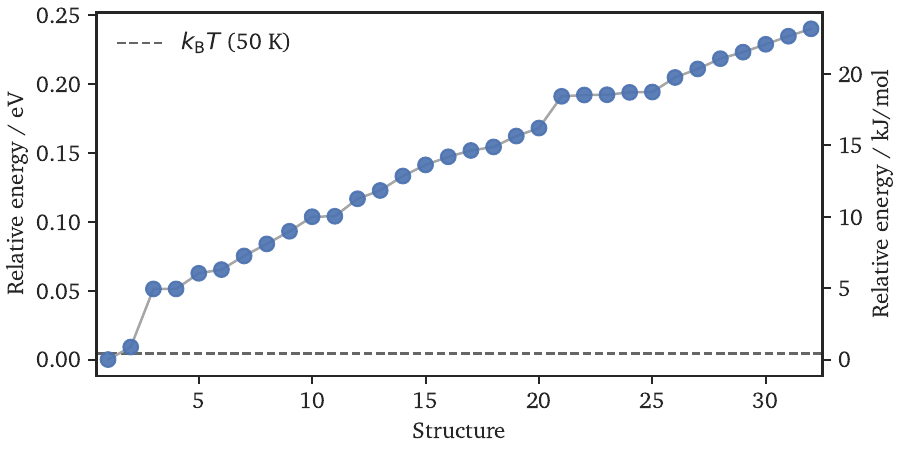}
    \caption{Reoptimized structures obtained from the CREST conformational search. The 32 final structures are shown together with the thermal energy limit corresponding to 50~K. Under these conditions, only the first and second structures are expected to be significantly populated. Calculations were performed at the $\omega$B97X-D3/aug-cc-pVDZ level of theory.}
    \label{fig:orca_opt}
\end{figure}

After reoptimization with DFT based on $\omega$B97X-D3/aug-cc-pVDZ using ORCA~6.0.1,\cite{neeseSoftwareUpdateORCA2025} the structures were screened again for duplicates, yielding 32 final structures within an energy window of 0.25~eV, as shown in Fig.~\ref{fig:orca_opt}. Two nearly isoenergetic minimum-energy structures were identified, separated by an energy difference of 0.01~eV. The next higher-lying structure was found to be 0.05~eV above these two minima. Considering the estimated molecular beam temperature of approximately 50~K, the populated dimer structures are expected to be dominated by these two structural motifs.

Frequency calculations were performed to verify that the optimized geometries correspond to true minima on the potential energy surface. To obtain a more accurate description of the dimers, the DFT integration grid \texttt{DefGrid3} was used for all ORCA calculations.

Subsequently, excited-state geometry optimizations for the S$_1$ and S$_2$ states were carried out for all 32 structures. For the S$_2$ state, the structure corresponding to the most stable ground-state geometry also remained the lowest-energy structure after excited-state optimization. In contrast, for the S$_1$ state, this structure was found to lie higher in energy than several other optimized geometries.

A notable challenge in the excited-state optimizations arises from the very shallow nature of the dimer potential energy surface. Although the optimizations formally converged using tightened SCF convergence criteria, \texttt{TightSCF}, and tightened geometry optimization criteria, \texttt{TightOpt}, subsequent frequency calculations still revealed small residual supramolecular vibrational modes. This indicates that the optimized excited-state structures are located on a very flat region of the potential energy surface, where weak intermolecular degrees of freedom remain difficult to converge completely.
\clearpage

\subsection{TDDFT calculations for the dimer structures}

\begin{figure}[h]
    \centering
    \includegraphics[]{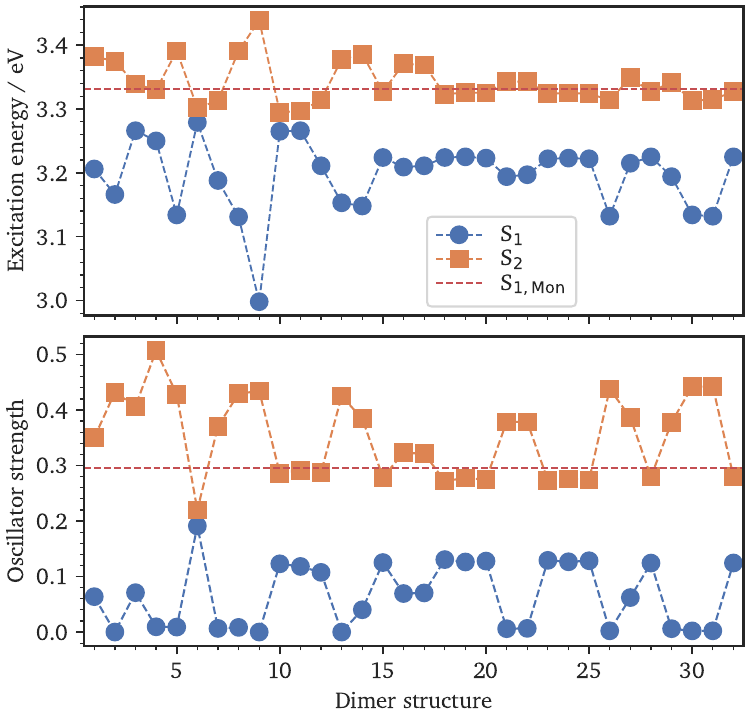}
    \caption{Calculated excitation energies and oscillator strengths of the investigated dimer structures. The dimer geometries were optimized in the electronic ground state using ORCA, as shown in Fig. \ref{fig:orca_opt}. TDDFT calculations were subsequently performed on the optimized S$_0$ structures. The vertical excitation energies of the S$_1$ and S$_2$ states and the corresponding oscillator strengths are shown as a function of the dimer structure. The dashed reference line indicates the corresponding S$_1$ value of the monomer. For all investigated structures, the S$_2$ state exhibits the larger oscillator strength and therefore represents the brighter excitonic state, which is indicative of H-aggregate type coupling.}
    \label{fig:dimer_tddft}
\end{figure}
\clearpage
\subsection{Cation dimer}

\begin{figure}[h]
    \centering
    \includegraphics[]{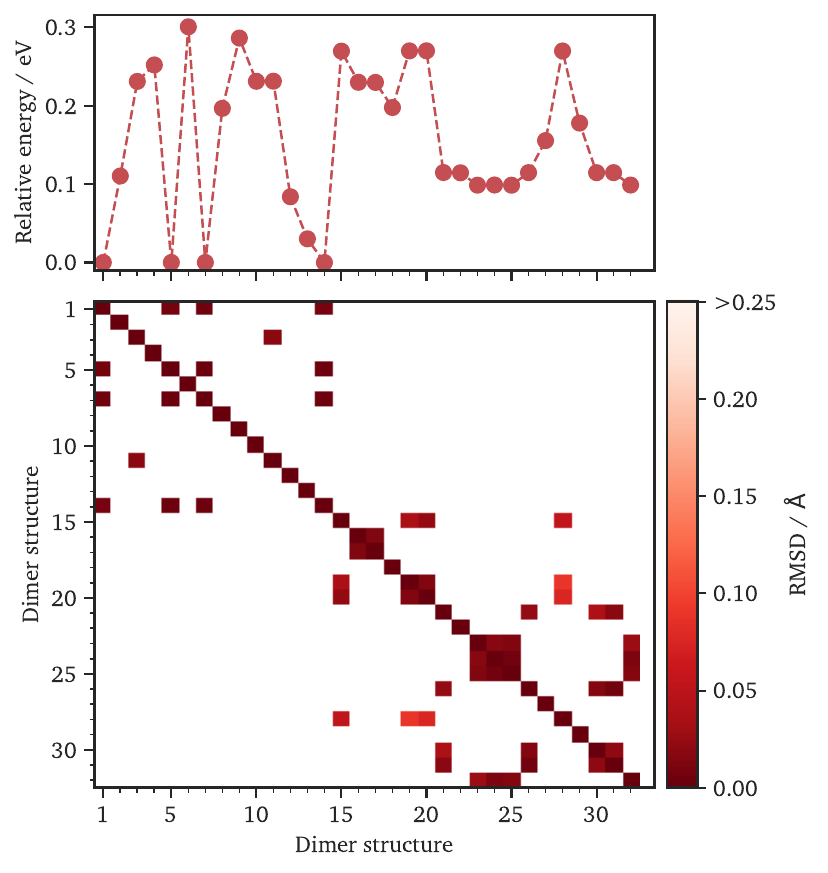}
    \caption{Relative energies and structural similarity of optimized cationic dimer structures obtained from the optimized neutral ground-state geometries. The upper panel shows the relative energies of the converged cation structures, referenced to the lowest-energy cation minimum. The lower panel shows the pairwise RMSD matrix used to assess whether the optimized geometries converge to the same cationic minimum. Structures 1, 5, 7, and 14 correspond to the same displaced minimum-energy cation structure reported in the main paper.}
    \label{fig:Cation_opt}
\end{figure}

\clearpage

\subsection{Dissociation limit}
The dissociation energy of the cationic dimer state was calculated including
zero-point energy corrections according to

\begin{equation}
E_{\mathrm{diss}}^{\mathrm{ZPE}}(i)
=
\left(E_{\mathrm{mon}}^{S_0}+\mathrm{ZPE}_{\mathrm{mon}}^{S_0}\right)
+
\left(E_{\mathrm{mon}}^{D_0}+\mathrm{ZPE}_{\mathrm{mon}}^{D_0}\right)
-
\left(E_{\mathrm{dim}}^{D_0}(i)+\mathrm{ZPE}_{\mathrm{dim}}^{D_0}(i)\right).
\label{eq:e_diss}
\end{equation}

Here, \(E_{\mathrm{mon}}^{S_0}\), \(E_{\mathrm{mon}}^{D_0}\), and
\(E_{\mathrm{dim}}^{D_0}(i)\) denote the electronic energies of the neutral
monomer, the cationic monomer, and the cationic dimer state \(i\),
respectively. The corresponding \(\mathrm{ZPE}\) terms denote the harmonic
zero-point energy corrections. The dissociation limit therefore consists of
one neutral and one cationic monomer.

The basis-set superposition error (BSSE) was corrected using the
Boys--Bernardi counterpoise method.\cite{boysCalculationSmallMolecular1970} For this purpose, each monomer was
calculated at its geometry within the optimized dimer while retaining the
basis functions of the second monomer as ghost functions. Since the system is
a homodimer, both charge assignments, \(A^{+}+B\) and \(A+B^{+}\), were
evaluated and their BSSE corrections were averaged:

\begin{equation}
\overline{\delta}_{\mathrm{BSSE}}(i)
=
\frac{1}{2}
\left[
\delta_{\mathrm{BSSE}}^{A^{+}B}(i)
+
\delta_{\mathrm{BSSE}}^{AB^{+}}(i)
\right].
\end{equation}

The counterpoise- and ZPE-corrected dissociation energy was then obtained as

\begin{equation}
E_{\mathrm{diss}}^{\mathrm{CP+ZPE}}(i)
=
E_{\mathrm{diss}}^{\mathrm{ZPE}}(i)
-
\overline{\delta}_{\mathrm{BSSE}}(i).
\label{eq:e_diss_cp}
\end{equation}

\begin{table}[htbp]
\centering
\caption{Adiabatic dissociation energies of the lowest-energy cationic \textbf{1} dimer structure. $E_{\mathrm{diss}}$ denotes the electronic dissociation energy and $E_{\mathrm{diss}}^{\mathrm{ZPE}}$ includes the harmonic zero-point energy. The counterpoise-corrected dissociation energy, $E_{\mathrm{diss}}^{\mathrm{CP+ZPE}}$, is obtained according to Eq.~\ref{eq:e_diss_cp}.}
\label{tab:cation-counterpoise-best-ev}
\setlength{\tabcolsep}{7pt}
\renewcommand{\arraystretch}{1.15}
\begin{tabular}{c c c c c c}
\toprule
Electronic state & $E_\text{diss}$/eV & $E^{\text{ZPE}}_\text{diss}$/eV & $\overline{\delta}_{\mathrm{BSSE}}$/eV & $E_{\mathrm{diss}}^{\mathrm{CP}}$/eV & $E_{\mathrm{diss}}^{\mathrm{CP+ZPE}}$/eV \\
\midrule
D$_0$ & 1.26 & 1.24 & 0.14 & 1.12 & 1.10 \\
\bottomrule
\end{tabular}
\end{table}

\begin{figure}[h]
    \centering
    \includegraphics[]{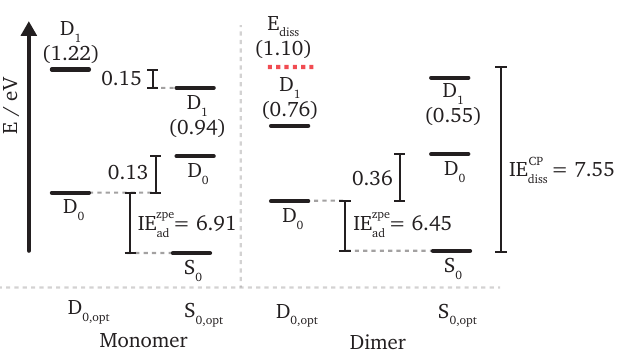}
    \caption{Adiabatic energy diagram of monomeric \textbf{1} (left) and the lowest-energy dimer structure (right). The cationic ground and first excited doublet states, D$_0$ and D$_1$, are shown at the optimized D$_{0,\mathrm{opt}}$ (see Fig. \ref{fig:Cation_opt}) and S$_{0,\mathrm{opt}}$ geometries.  Values in parentheses denote the corresponding vertical D$_1$ excitation energies. The red dashed line marks the adiabatic dissociation energy of the cationic dimer, $E_{\mathrm{diss}}^{\mathrm{CP+ZPE}} = 1.10$ eV, including counterpoise and zero-point-energy corrections. Zero-point-energy-corrected adiabatic ionization energies, $\mathrm{IE}_{\mathrm{ad}}^{\mathrm{ZPE}}$, and the counterpoise-corrected dissociative ionization energy, $\mathrm{IE}_{\mathrm{diss}}^{\mathrm{CP}}$, are also indicated. Calculated according to Eq.~\ref{eq:e_diss_cp}; the resulting values are summarized in Tab.~\ref{tab:cation-counterpoise-best-ev}.
}
    \label{fig:energy_cation}
\end{figure}

\clearpage

\subsection{NEB minimum-energy pathway}

\begin{figure}[h]
    \centering
    \includegraphics[]{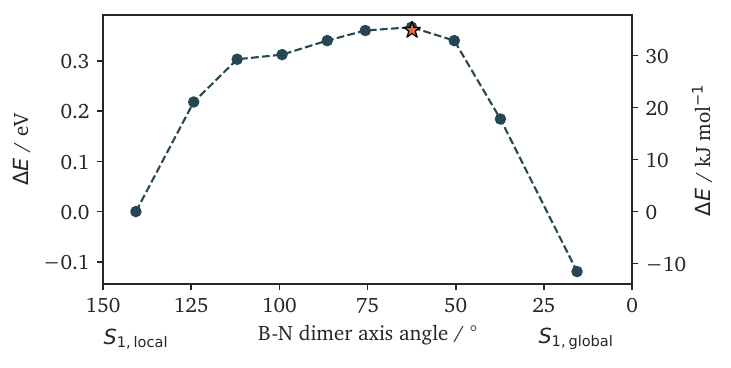}
    \caption{Nudged elastic band pathway connecting the local and global S$_1$ minima of the \ce{BN} dimer. The local S$_1$ minimum is located on the left-hand side of the reaction path at a \ce{BN} dimer axis angle of 140.7$^\circ$, while the global S$_1$ minimum is reached on the right-hand side at 15.6$^\circ$. All energies are referenced to the local S$_1$ minimum. The local-to-global S$_1$ barrier amounts to 0.36 eV and is associated with the relative rotation of the $\pi$-planes. The star marks the optimised transition state. The \ce{BN} dimer axis angle defines the relative orientation of the two \ce{BN} units, where 0$^\circ$ corresponds to an aligned \ce{BN-BN} configuration. The pathway was calculated using excited-state NEB-TS optimisation at the $\omega$B97X-D3/aug-cc-pVDZ level of theory without the Tamm–Dancoff approximation.\cite{asgeirssonNudgedElasticBand2021} Eight NEB images were generated by image-dependent pair potential interpolation without prior optimisation, employing tight SCF convergence and the DefGrid3 integration grid.}
    \label{fig:NEB_path}
\end{figure}

\subsection{Dimer structure - Surface hopping}

For trajectory-based nonadiabatic dynamics simulations, Tully's fewest-switches surface hopping\cite{tullyMolecularDynamicsElectronic1990} procedure was applied in combination with a local diabatization scheme\cite{granucciDirectSemiclassicalSimulation2001,plasserSurfaceHoppingDynamics2012} using TD-DFT at the $\omega$B97X-D/cc-pVDZ level of theory.\cite{dunningGaussianBasisSets1989,chaiLongrangeCorrectedHybrid2008} The electronic structure calculations were carried out with the Q-Chem software package.\cite{epifanovskySoftwareFrontiersQuantum2021} Starting from the S$_0$-optimized lowest-energy dimer structure, a total of 50 initial structures were generated by sampling a harmonic canonical Wigner distribution at 50 K and were subsequently relaxed by propagating them for 100 fs in the electronic ground state without a thermostat.\cite{wignerQuantumCorrectionThermodynamic1932,bonacic-kouteckyTheoreticalExplorationUltrafast2005} The geometries were launched from the S$_2$ electronic state, and the trajectories were propagated for 100 fs with a time step of 0.25 fs. Within this framework, the nuclei are propagated classically on the active electronic state using the Verlet algorithm, whereas the electronic amplitudes are evolved quantum mechanically along the nuclear trajectory. The coupling analysis was restricted to the nonadiabatic coupling between the S$_2$ and S$_1$ states in order to target the evolution of the initially excited S$_2$ state and possible short-time population transfer to S$_1$. Accordingly, the simulations focus on the S$_2$~$\rightarrow$~S$_1$ internal conversion pathway and provide insight into the early excited-state dynamics, while subsequent relaxation pathways involving the electronic ground state are not explicitly resolved. The explicit evaluation of couplings to S$_0$ was omitted because of the considerably larger computational cost. Due to the high computational demand, with an average wall time of approximately 70 min per propagation step, the total propagation time was limited to 100 fs. The basis was not reduced further, as an overly small basis would lead to an unreliable description of the electronically excited states in the investigated \ce{BN}-doped system. The calculations were performed in parallel on 20 processors using Intel(R) Xeon(R) CPU E5-2680 v4 @ 2.40GHz hardware.

\subsection{Geometric descriptors for dimer plane alignment and curvature}

The monomer planes were fitted independently for each dimer structure using all non-hydrogen atoms of the respective monomer. For monomer \(m\), with heavy-atom coordinates \(\mathbf{r}_{i,m}\), the centroid is
\[
\mathbf{c}_m = \frac{1}{N_m}\sum_i \mathbf{r}_{i,m}.
\]
The fitted plane \(p_m\) is defined by the unit normal vector \(\mathbf{n}_m\), obtained as the right-singular vector corresponding to the smallest singular value of the centered coordinate matrix \(\mathbf{r}_{i,m}-\mathbf{c}_m\). Thus,
\[
p_m:\quad \mathbf{n}_m \cdot (\mathbf{r}-\mathbf{c}_m)=0.
\]
The plane--plane distance was calculated from the centroid displacement projected onto both fitted plane normals,
\[
d_{\mathrm{p}\text{-}\mathrm{p}}
=
\frac{1}{2}
\left(
\left|(\mathbf{c}_2-\mathbf{c}_1)\cdot\mathbf{n}_1\right|
+
\left|(\mathbf{c}_2-\mathbf{c}_1)\cdot\mathbf{n}_2\right|
\right).
\]
The distance of an atom \(X\) to the opposite monomer plane was computed as the absolute point-to-plane distance,
\[
d_{X_m\text{-}\mathrm{p}_{\bar{m}}}
=
\left|
(\mathbf{r}_{X,m}-\mathbf{c}_{\bar{m}})\cdot\mathbf{n}_{\bar{m}}
\right|,
\]
and the reported mean projection distance is
\[
\bar{d}_{X\text{-}\mathrm{p}}
=
\frac{1}{2}
\left(
d_{X_1\text{-}\mathrm{p}_2}
+
d_{X_2\text{-}\mathrm{p}_1}
\right).
\]
For BN dimers, \(X\) corresponds to N or B; for the phenanthrene reference dimers, \(X\) corresponds to the carbon atoms at the respective N and B substitution positions, denoted C(N) and C(B). The interplane tilt angle was calculated from the absolute scalar product of the two plane normals,
\[
\alpha_{\mathrm{p}}
=
\cos^{-1}\left(|\mathbf{n}_1\cdot\mathbf{n}_2|\right).
\]
To quantify monomer curvature, the heavy atoms of each monomer were expressed in the local plane coordinate system \((x_i,y_i,z_i)\), where \(z_i\) is the displacement along \(\mathbf{n}_m\). A quadratic surface
\[
z = ax^2 + by^2 + cxy + dx + ey + f
\]
was fitted by least squares. The corresponding curvature matrix is
\[
\mathbf{H}
=
\begin{pmatrix}
2a & c \\
c & 2b
\end{pmatrix},
\]
with principal curvatures \(\kappa_1\) and \(\kappa_2\) given by the eigenvalues of \(\mathbf{H}\). The reported curvature descriptor is the root-mean-square principal curvature,
\[
\kappa_{\mathrm{rms}}
=
\sqrt{\frac{\kappa_1^2+\kappa_2^2}{2}}.
\]

\subsection{Phenanthrene dimer}

\begin{table}[htbp]
\centering
\caption{Distances, curvature, and interplane angle for the lowest-energy phenanthrene reference dimer structures; p denotes the fitted monomer plane.}
\label{tab:phenanthrene-lowest-energy-plane-curvature-tilt}
\begin{tabular}{cccccc}
\toprule
\shortstack{Electronic\\state} & \raisebox{0.5\normalbaselineskip}{$d_{\mathrm{p}\text{-}\mathrm{p}}$/\text{\AA}} & \raisebox{0.5\normalbaselineskip}{$\bar{d}_{\mathrm{C(N)}\text{-}\mathrm{p}}$/\text{\AA}} & \raisebox{0.5\normalbaselineskip}{$\bar{d}_{\mathrm{C(B)}\text{-}\mathrm{p}}$/\text{\AA}} & \raisebox{0.5\normalbaselineskip}{$\kappa_{\mathrm{rms}}$/\text{\AA}$^{-1}$} & \raisebox{0.5\normalbaselineskip}{$\alpha_{\mathrm{p}}$/deg} \\
\midrule
S$_0$ & 3.44 & 3.44 & 3.38 & 0.0077 & 5.58 \\
S$_1$ & 3.24 & 3.29 & 3.29 & 0.0154 & 8.83 \\
\bottomrule
\end{tabular}
\end{table}

\begin{figure}[h]
    \centering
    \includegraphics[]{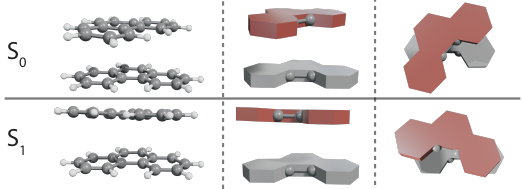}
    \caption{Optimized phenanthrene dimer structures of the S$_0$, S$_1$ states. Simplified side and bird's-eye-view representations are included to illustrate the relative orientation of the dimer units.}
    \label{fig:CC_all}
\end{figure}

\begin{figure}
    \centering
    \includegraphics[]{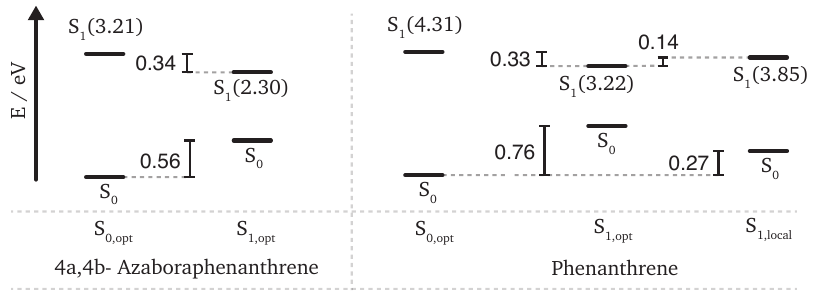}
    \caption{Ground-state energy differences between the minimum S\(_0\)-optimized and S\(_1\)-optimized geometries of \textbf{1} and phenanthrene dimer. For phenanthrene dimer, the S\(_{1,\mathrm{local}}\) structure, which most closely resembles the S\(_1\)-optimized geometry of \textbf{1} dimer, is additionally shown. This local S\(_1\) minimum lies \SI{0.14}{eV} above the global S\(_1\) minimum of phenanthrene. Values in parentheses denote the corresponding S\(_1\) excitation energies. All energies were calculated at the \(\omega\)B97X-D3/aug-cc-pVDZ level of theory.}
    \label{fig:energy_cc}
\end{figure}

\clearpage
\section{NMR spectra} \label{sec:nmr}
\begin{figure}[h]
    \centering
    \includegraphics[]{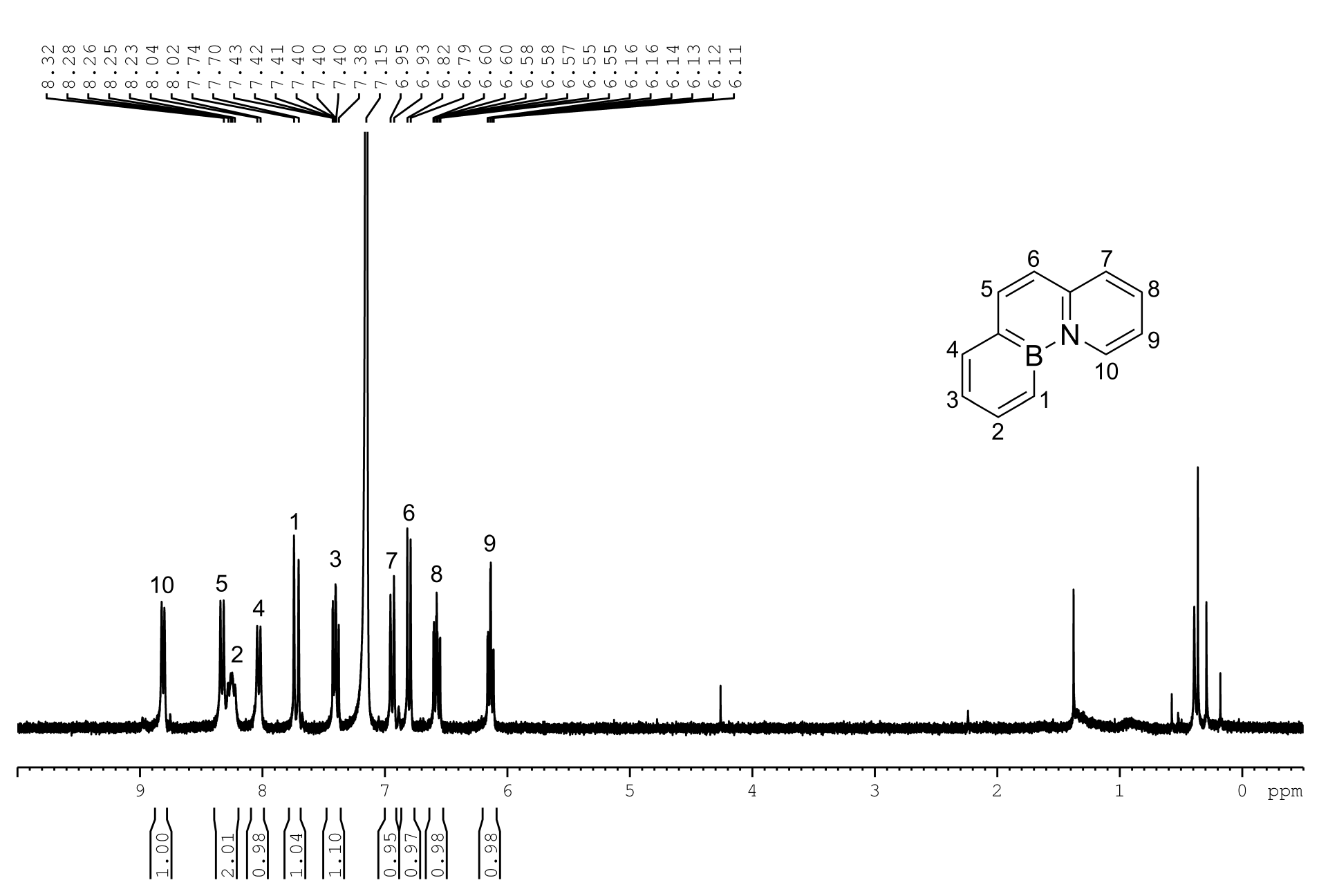}
    \caption{$^1$H-NMR (300 MHz) of compound \textbf{1} in C$_6$D$_6$.}
    \label{fig:1HNMR}
\end{figure}

\begin{figure}[h]
    \centering
    \includegraphics[]{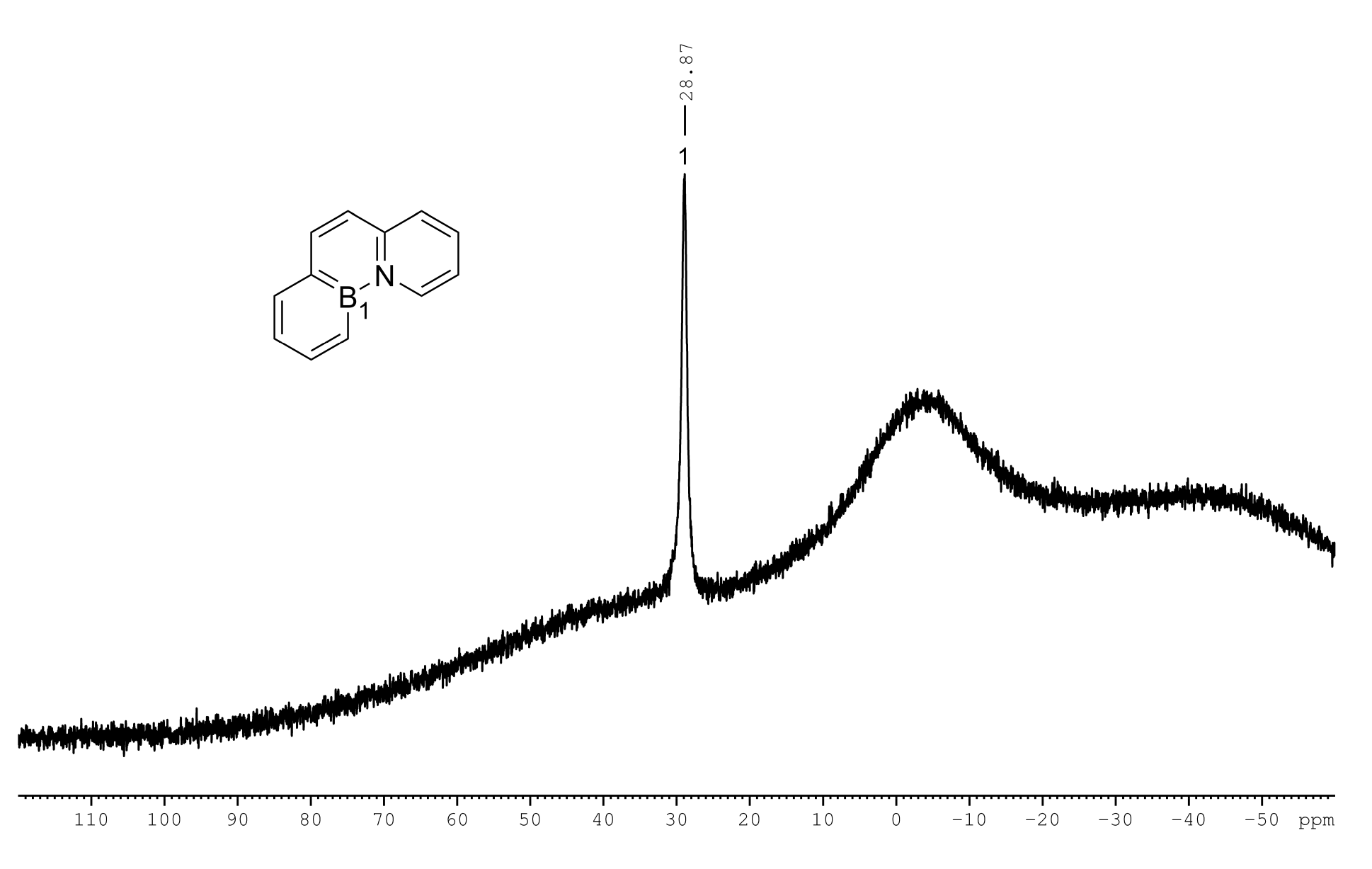}
    \caption{$^{11}$B\{$^{1}$H\}-NMR (96 MHz) of compound \textbf{1} in C$_6$D$_6$.}
    \label{fig:11BNMR}
\end{figure}

\clearpage



\balance


\bibliography{BN_Phen} 
\bibliographystyle{rsc} 